\documentclass{aa}
\usepackage{psfig}
\usepackage{natbib}
\bibpunct{(}{)}{,}{a}{}{,} 
\usepackage{graphicx}

\def\la{\;
\raise0.3ex\hbox{$<$\kern-0.75em\raise-1.1ex\hbox{$\sim$}}\; }
\def\ga{\;
\raise0.3ex\hbox{$>$\kern-0.75em\raise-1.1ex\hbox{$\sim$}}\; }

\newcommand{\zabs}{$z_{\rm abs}\,$}
\newcommand{\zem}{$z_{\rm em}\,$}

\newcommand{\kms}{km~s$^{-1}\,$}
\newcommand{\cm}{cm$^{-2}\,$}
\newcommand{\cmm}{cm$^{-3}\,$}
\newcommand{\etal}{{\it et al.}}

\begin{document}

\title{
Metal-rich absorbers at high redshifts: abundance patterns\thanks{Based 
on observations obtained with the UVES
at the VLT Kueyen telescope (ESO, Paranal, Chile).} 
}
\subtitle{}
\author{S. A. Levshakov\inst{1,2}
\and
I. I. Agafonova\inst{2}
\and
P. Molaro\inst{1}
\and
D. Reimers\inst{3}
\and
J. L. Hou\inst{4}
}
\institute{
INAF-Osservatorio Astronomico di Trieste, Via G. B. Tiepolo 11,
34131 Trieste, Italy
\and
Ioffe Physical-Technical Institute, 
Polytekhnicheskaya Str. 26, 194021 St.~Petersburg, Russia\\
\email{lev@astro.ioffe.rssi.ru} 
\and
Hamburger Sternwarte, Universit\"at Hamburg,
Gojenbergsweg 112, D-21029 Hamburg, Germany
\and
Key Lab. for Research in Galaxies and Cosmology, 
Shanghai Astronomical Observatory, CAS, 80 Nandan Road, Shanghai 200030, 
P.R. China
}
\date{Received 00  ; Accepted 00}
\abstract
{}
{To study chemical composition of metal-rich absorbers at high redshifts
in order to understand their nature and to determine 
sources of their metal enrichment. 
}
{From six spectra of high-$z$ QSOs, we select eleven 
metal-rich, $Z \ga Z_\odot$, and optically-thin  
to the ionizing radiation, 
$N$(\ion{H}{i}) $< 10^{17}$ \cm,
absorption systems ranging between $z=1.5$ and $z=2.9$ and
revealing lines of different ions in subsequent
ionization stages. Computations are performed 
using the Monte Carlo Inversion (MCI) procedure complemented with
the adjustment of the spectral shape of the ionizing radiation. 
This procedure along with selection criteria 
for the absorption systems guarantee the accuracy of 
the ionization corrections and of
the derived element abundances (C, N, O, Mg, Al, Si, Fe).  
}
{The majority of the systems (10 from 11) 
show abundance patterns which 
relate them to outflows from low and
intermediate mass stars. One  absorber is enriched prevalently 
by SNe~II, however,
a low percentage of such systems in our sample is conditioned by 
the selection criteria. 
All systems have sub-kpc linear sizes along the line-of-sight 
with many less than $\sim 20$ pc. 
In several systems, silicon is deficient, 
presumably due to the depletion
onto dust grains in the envelopes of dust-forming stars and the
subsequent gas-dust separation. 
At any value of [C/H], nitrogen can be either
deficient, [N/C] $< 0$, or enhanced, [N/C] $> 0$, 
which supposes that the nitrogen enrichment occurs irregularly.
In some cases, 
the lines of \ion{Mg}{ii} $\lambda\lambda2796, 2803$ 
appear to be shifted, probably as a result of an enhanced
content of heavy isotopes $^{25}$Mg and $^{26}$Mg in the absorbing gas 
relative to the solar isotopic composition. 
Seven absorbers are characterized by low mean ionization parameter $U$, 
$\log U < -2.3$,
among them only one system has a redshift $z > 2$ 
(\zabs = 2.5745) whereas all others
are found at $z  \sim 1.8$.
This statistics is not affected by any selection criteria and reflects the
real rise in number of such systems at $z < 2.0$. 
Comparing the space number density of metal-rich absorbers with
the comoving density of star-forming galaxies at $z \sim 2$,
we estimate that the circumgalactic volume of each galaxy is populated
by $\sim 10^7-10^8$ such absorbers 
with total mass $\la 1/100th$ of the stellar galactic mass. 
Possible effects of high metal
content on the peak values of star-forming and AGN activities at $z \sim 2$
are discussed.  
}
{}
\authorrunning{S. A. Levshakov \etal}
\titlerunning{Metal-rich absorbers at high redshifts: abundance patterns}

\maketitle

\section{Introduction}
\label{sect-1}

The majority of metal-bearing absorbers detected in quasar 
spectra demonstrate very
low metallicity of about 1/30 of the solar value, $Z_\odot$, and below
(Songaila \& Cowie 1996; Simcoe \etal\ 2004). 
Compared to the metal-poor
absorbers, the systems with high metallicity~--- from solar to oversolar~---
occur in quasar spectra quite seldom. 
A particular interest to the metal-rich absorbers stems from the fact that 
high metallicities are usually measured in the central parts of galaxies
(Rich \etal\ 2007; Cohen \etal\ 2008; Zoccali \etal\ 2008; Davies \etal\ 2009)
and are closely related to the starburst activity and chemical evolution of 
AGNs/QSOs (Hamann \& Ferland 1999; D'Odorico \etal\ 2004;
Polletta \etal\ 2008; Silverman \etal\ 2009). 
At high redshifts, the only way to gain an insight into the chemical 
composition of gas distributed over large cosmological distances
is through the analysis of absorption systems in 
quasar spectra. Individual element abundances, or abundance patterns, 
derived from the metal-rich systems can be used to reconstruct
their chemical history and to estimate  relative 
contributions from SNe~II, SNe~Ia, and AGB-stars.

The luminous quasar space density distribution (Fan \etal\ 2001)
and observations of high-redshift galaxies (Dickinson \etal\ 2003)
show that the redshift $z \sim 2$ represents a turning point
in the  global cosmic evolution. 
The study of the chemical enrichment at
different $z$ can lead to important clues concerning the factors which
cause the peak values of AGN and star-forming activities at $z \sim 2$.

To obtain chemical enrichment patterns of the absorbing gas,
accurate abundances of different elements are required. 
Abundance measurements in high-metallicity damped Ly-$\alpha$
absorbers (DLAs) and sub-DLAs 
where ionization corrections are supposed to be small 
are limited either to low abundant species like Zn, Mn, and Cr, 
or to those which have transitions with low oscillator strengths like 
Fe and Si since heavily saturated lines of other elements prevent their 
accurate column density determination 
(Prochaska \etal\ 2006; P\'eroux \etal\ 2006; Meiring \etal\ 2007;
Meiring \etal\ 2008; P\'eroux \etal\ 2008). 
However, just Fe and Si (as a proxy for $\alpha$-elements) 
which belong to the key  species in the abundance pattern, 
can be severely depleted 
(Hou \etal\ 2001; Vladilo 2002; Vladilo \& P\'eroux 2005; Quast \etal\ 2008), 
thus hampering significantly the identification of the enrichment agents.
Note that metal-rich DLAs are very rare~--- up to now, only about 
a dozen is detected. 

On the other hand,
in absorbers which are optically-thin to the ionizing radiation\footnote{Column 
density of neutral 
hydrogen $N$(\ion{H}{i}) $< 3\times10^{17}$ \cm. }
the ionization correction factors  are large 
and, consequently, the resulting element abundances are very sensitive 
to their values. 
In general, ionization corrections are determined by the ionization
parameter $U$ {\it and} by the spectral shape of the ionizing radiation.
To increase the reliability of the measured abundances both 
a special selection of absorption systems and an adequate mathematical treatment 
of the inverse spectroscopy problem are required. 

An approach employed in the present paper is based on the physical model 
of absorbing gas which assumes fluctuating gas density and velocity fields 
inside the absorber. 
This model is a generalization of a commonly used
approximation of a plane-parallel gas slab of uniform density and 
a microturbulent treatment of the velocity field. 
The corresponding computational procedure used to 
calculate the metal abundances includes also the adjustment 
of the spectral shape of the ionizing radiation 
(see Sect.~\ref{sect-2}). 

Since relative element ratios can differ from the solar pattern, it should be
clarified on base of which species we conclude whether the gas is
metal-rich or not. Metallicity in stars is usually characterized by the
iron content, that of strong absorption systems (LLSs and DLAs)~---
by the content of zinc or sulfur which are not condensed 
significantly into dust.
In optically-thin absorption systems, metallicity is
conventionally estimated on base of carbon.
Thus, in the present metal-rich sample we included  absorbers with
carbon abundances near and above the solar value,
(C/H)$_\odot = 2.45\times10^{-4}$ (Asplund \etal\ 2004).

The selection of systems occurred in the following way. In the framework
of our project to study the spectral energy distribution (SED)
of the ionizing radiation the quasar spectra were searched for the
optically-thin absorption systems revealing lines of different ions 
(Agafonova \etal\ 2005). 
The presence of at least one pair of ions of the same element 
(e.g. \ion{C}{ii}, \ion{C}{iv}) 
was essential for the system to be included in the sample. 
These subsequent ions ensure that the mean ionization parameter can be 
estimated without any additional assumptions.
In total, about 50 absorbers identified in high-resolution spectra
of $\sim20$ QSOs with emission redshifts $1.7 < z < 4.5$ 
were selected and analyzed. 
The derived metallicities range from 1/1000 to several times solar values.
A part of results was described in our publications dealing with
the evolution of the background SED with redshift 
(Reimers \etal\ 2006; Agafonova \etal\ 2007) and the restoration of
the SED of the outcoming quasar radiation on base of
associated systems (Reimers \etal\ 2005; 
Levshakov \etal\ 2008, hereafter Paper~I). 
These publications included already several systems with high metal contents,
however, enrichment scenarios were not elaborated. This is just
what the present paper is focused on where we consider additional 7 metal-rich
absorbers. With the previously described systems, 
the sample of metal-rich absorbers consists now of 
11 systems with redshifts ranging
from $z = 1.5$ to $z = 2.9$.

\begin{figure*}[t]
\vspace{0.0cm}
\hspace{-0.2cm}\psfig{figure=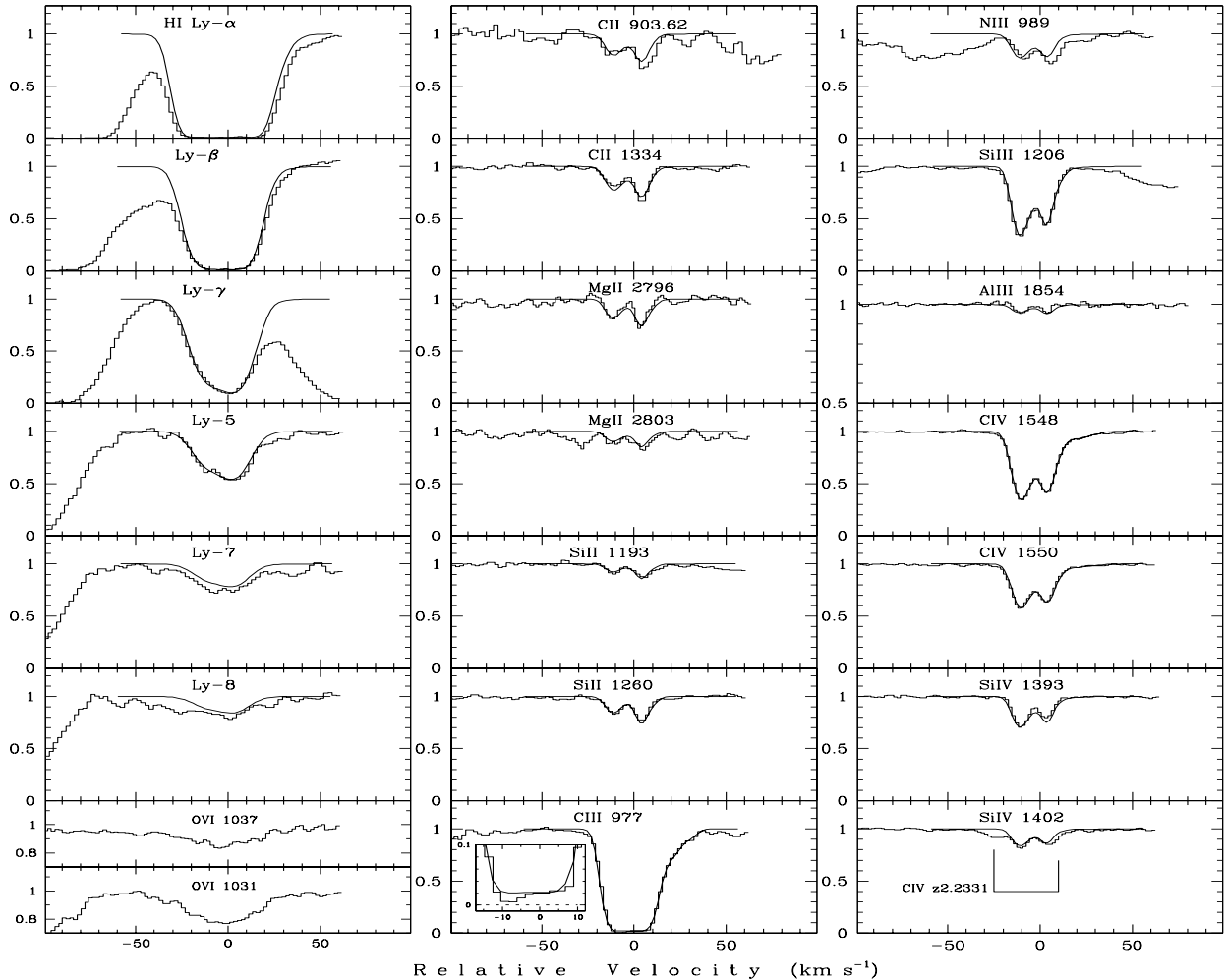,height=14cm,width=18cm}
\vspace{-0.5cm}
\caption[]{
Hydrogen and metal absorption lines from the \zabs = 2.5745 system
towards \object{HE1347--2457} (solid-line histograms).
The vertical axis is normalized intensities.
The zero radial velocity is fixed at $z = 2.5745$.
Synthetic profiles are plotted by the smooth curves. 
The wide profiles of the \ion{O}{vi}$ \lambda\lambda 1031, 1037$ 
lines show that the narrower line metal-rich system is embedded in  
a hot gas (seen also in \ion{H}{i} Ly-$\alpha$ and 
Ly-$\beta$ lines as a difference between
the observed and synthetic profiles). 
The insert in panel \ion{C}{iii} $\lambda977$ demonstrates 
a non-zero residual intensity at the line center.
The identified blend is marked in panel \ion{Si}{iv} $\lambda1402$.
}
\label{fg_1}
\end{figure*}

\section{Computational method}
\label{sect-2}

Absorption systems are analyzed by means of the Monte Carlo Inversion 
(MCI) procedure
described in Levshakov, Agafonova, \& Kegel (2000, hereafter LAK),
with further details in Levshakov \etal\ (2003). 
This procedure supposes that all lines observed in a metal system 
are formed in the gas slab with fluctuating density 
$n_{\scriptscriptstyle \rm H}(x)$ and velocity $v(x)$ fields 
(here $x$ is the space coordinate
along the line-of-sight within the absorber).
Further assumptions are that within the absorber the metal abundances are
constant and the gas is in thermal and ionization equilibrium. 
The procedure is applicable for optically-thin systems. 
The inputs are the observed line profiles and 
the ionization curves for each ion (hydrogen + metals) included in the analysis. 
The ionization curves (the dependence of the ion fraction
on the ionization parameter $U$) are
computed with the photoionization code CLOUDY version 07.02.01
(last described by Ferland \etal\ 1998) which in turn uses as the inputs
an ionizing continuum and a set of element abundances. 
The fitting parameters (outputs) of the MCI procedure are: 
the mean ionization parameter $U_0$, 
the total hydrogen column density $N_{\scriptscriptstyle \rm H}$, 
the line-of-sight velocity dispersion, $\sigma_v$, and
the density dispersion, $\sigma_y$, of the bulk material 
[$y \equiv n_{\scriptscriptstyle \rm H}(x)/n_0$], 
and the contents $Z_a$ of metals included in the objective function 
[Eqs.(29, 30) in LAK].
The fitting parameters are estimated by minimization of residuals between 
synthetic and observed line profiles.
As in all Monte Carlo methods, integrals [here the intensity at a point 
within the line profile, see Eq.(12) in LAK] are calculated by modeling 
the distribution of the integrated function. This function is a convolution 
of the gas density 
and the local absorption coefficient (which depends on the velocity
through the irregular Doppler shifts) 
and, hence, the integrated function 
is realized through the density and velocity 
distributions over the integration path. Both the
distributions of  $n_{\scriptscriptstyle \rm H}(x)$ and $v(x)$ 
are represented by their sampled values $\{n_i\}$ and $\{v_i\}$ 
at equally spaced points $x_i$ along the line-of-sight,
and optimal configurations of $\{n_i\}$ and $\{v_i\}$ are estimated by
the simulated annealing algorithm. 
With the estimated values of the fitting parameters and
the distributions of $n_{\scriptscriptstyle \rm H}(x)$ and $v(x)$, 
we can calculate the column densities of all ions [Eqs.(24, 25) in LAK] and the
mean values for the kinetic temperature,
$T_{\rm kin}$, the gas density, $n_0$, and the linear size, $L$,
of the absorber along the line-of-sight (see below).

If the metal abundances $Z_a$ obtained from the MCI procedure differ 
from those used in CLOUDY to calculate the ionization curves, then 
these curves are re-calculated with the updated abundances and the 
MCI procedure is run again. The iterations stop after the concordance 
between the abundances used in CLOUDY and derived from MCI is reached.

At this stage of calculations, the shape (SED) of the ionizing radiation 
(which comes into the MCI procedure through the ionization curves) 
is treated as an external parameter, i.e., it is keeping constant. 
In case when a first guessed SED cannot provide an acceptable solution,
a complemented computational routine aimed at adjusting the spectral shape 
is applied. The corresponding algorithm based on the response
surface methodology from the theory of experimental design
is described in detail in Agafonova \etal\ (2005, 2007). 
It includes a parameterization of the SED 
by means of a set of factors and the estimation of their  values in a way
to ensure that all obtained physical parameters are self-consistent and 
all observed profiles are described with a sufficient accuracy 
(i.e., $\chi^2 \sim 1$). 

It is known from both model calculations  
(Haardt \& Madau 1996; Fardal \etal\ 1998; Madau \& Haardt 2009) 
and from SEDs restored from observational data (Agafonova \etal\ 2005, 2007) 
that spectral shapes of the intergalactic
ionizing background in the energy range $E > 1$ Ryd 
have complex forms with an emission bump at $E \la 3$ Ryd and
a stepwise break at $E > 3$ Ryd.  
However, as noted in Agafonova \etal\ (2007; Sect. 3.1.2), 
at high metallicities the dependence of the ion fractions on the 
absolute abundances (especially on the contents of the
most abundant elements such as carbon and oxygen)
is comparable or even overrides the dependence on the spectral 
shape.
For the systems considered in the present study this means 
that the inherent uncertainty in the values of
the absolute element abundances makes it impossible to estimate 
the SED of the underlying ionizing continuum in full detail. 
That is why the spectral shape is represented and estimated 
in a simplified form including only 2 factors: 
a power law index $\alpha$ ($\nu^{-\alpha}$) in the range $1 < E < 4$ Ryd, 
and a depth of a break at 4 Ryd. The power law index for the energy
range beyond this break ($E > 4$ Ryd) was fixed as $1 - 0.8\alpha$.

The mean linear size $L$ of an absorbing cloud along the line of sight 
can be estimated as
$L = N_{\scriptscriptstyle \rm H}/n_0$, 
where $N_{\scriptscriptstyle \rm H}$ is the total hydrogen column density, 
and $n_0$ is the mean gas density of the absorber. 
The mean gas density is calculated from the mean ionization parameter $U_0$ 
and the density dispersion $\sigma_y$ (LAK):
\begin{equation}
n_0 = \frac{n_{\rm ph}}{U_0}\ {\rm e}^{\sigma^2_y}\ .
\label{S1-E1}
\end{equation}
Here
$n_{\rm ph}$ is the number density of photons (in units of \cmm)
with energies above 1 Ryd
\begin{equation}
n_{\rm ph} = \frac{4\pi}{c\ h} J_{912}
\int^\infty_{\nu_c} \left(
\frac{J_\nu}{J_{912}}\right) \frac{d\nu}{\nu}\, ,
\label{S1-E2}
\end{equation}
where $c, h, \nu_c$, and $J_\nu$ are the speed of light, the Planck constant,
the frequency of the hydrogen Lyman edge, and the specific
intensity (in ergs \cm s$^{-1}$ sr$^{-1}$ Hz$^{-1}$).
$J_{912}$ is the normalizing value and represents 
the intensity of the ionizing radiation at 
912 \AA\ (1 Ryd). As a reference, the value of $J_{912}$
for the intergalactic background at a corresponding redshift 
is taken from Haardt \& Madau (1996).    

{\rm 
If a strong radiation source such as a quasar is located close 
(up to $\sim$2 Mpc away) to the absorber, then the intensity $J_{912}$
of the incident ionizing radiation exceeds significantly the mean
intergalactic level (see, e.g., examples in Paper~I).
Thus, the gas density/linear size calculated on base of the
intergalactic $J_{912}$ represents in fact only a lower/upper limit,
respectively.

\begin{figure}[t]
\vspace{-1.0cm}
\hspace{-1.5cm}\psfig{figure=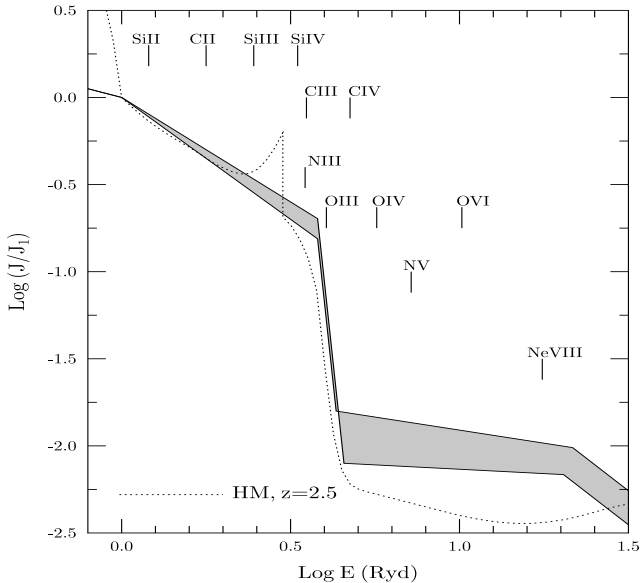,height=13cm,width=12cm}
\vspace{-4.5cm}
\caption[]{
The ionizing spectrum corresponding to the ionization 
state observed in the \zabs = 2.5745
system (\object{HE1347--2457}) is one from the shaded area.
For comparison the intergalactic ionizing spectrum at $z = 2.5$ modeled by   
Haardt \& Madau (1996) is shown by  the dotted line.
The spectra are normalized so that $J_\nu(h\nu =$ 1 Ryd) = 1.
The positions of ionization thresholds of different ions are
indicated by tick marks.
}
\label{fg_2}
\end{figure}

A few systems from the present study show signs of 
the incomplete coverage of the background light source. Computational
treatment of such systems occurred as described in Sect.~2 of Paper~I.  
The covering factor is assumed to be the same for the whole cloud,
i.e. it does not depend on
$\lambda$ (or, equivalently, on the radial velocity $v$).
However, different covering factors are allowed for different ions.
This follows from a model of the absorption 
arising in a gas cloud with fluctuating density. Namely, ions of
higher ionization stages trace rarefied gas which can be quite extended
(i.e., covers a greater part the background source)
whereas low ionization ions
originate in more dense and, hence, compact volumes
(i.e., the covering factor is smaller).  

The laboratory wavelengths and oscillator strengths for all lines except 
\ion{Si}{iii} were taken from Morton (2003). 
For \ion{Si}{iii} Morton gives 1206.500 \AA\
(see comments on page 190 in Morton 1991), however with this wavelength
the \ion{Si}{iii} line comes out shifted by $\sim$2.5 \kms\ 
relative to both \ion{Si}{ii} and \ion{Si}{iv} lines.
Since metal lines in absorption systems considered in the present paper 
are very narrow ($FWHM \la 15$ \kms) this shift becomes noticeable. 
On the other hand, other catalogs of atomic and ionic spectral lines
(e.g., Bashkin \& Stoner 1975; Kelly 1987) and
catalogs of solar lines (e.g., Sandlin \etal\ 1986; Curdt \etal\ 2001) give
for \ion{Si}{iii} the wavelength 1200.51 \AA. 
This value fits much better to what is
observed in the absorption systems and it was therefore 
adopted for the present work. 

Solar abundances were taken from Asplund \etal\ (2004). 
Note that their solar abundance of 
nitrogen, (N/H)$_\odot$ = $6\times10^{-5}$, is $\sim$1.4
times (0.15 dex) lower than that from Holweger (2001).

Quasar spectra were obtained with 
the UVES/VLT in the framework of the ESO Large
Program `QSO Absorption Line Systems' (ID No.166.A-0106).
Data reduction was performed by B. Aracil (Aracil \etal\ 2004).
Except \object{Q0329--385}, all QSOs used in the present paper 
have been discovered in course of the Hamburg/ESO survey 
(Wisotzki \etal\ 1996; Reimers \etal\ 1996; Wisotzki \etal\ 2000).

\begin{table*}[t!]
\centering
\caption{Physical parameters of metal-rich absorbers. }
\label{tbl-1}
\begin{tabular}{cllccccccc}
\hline
\hline
\noalign{\smallskip}
\# & \multicolumn{1}{c}{QSO} & \multicolumn{1}{c}{\zem} &
\zabs&$\log U_0$&$N$(\ion{H}), \cm&$n_0$, \cmm&$L$, pc &$\bar{T}_4^\dagger$ &
Ref.$^a$ \\
\noalign{\smallskip}
\hline 
\noalign{\medskip}
1&\object{HE1347--2457} & 2.578 &
2.5745 & -2.3 & 3.5E17 & $>$5E-3 &  $<$25 & 0.7& 1\\ 
2&&&1.7529 & -2.5 &(5 -- 8)E18 & 0.007 -- 0.1 & 12 -- 500 & 1.1&1\\
3&&&1.5080 & -2.5 & (5 -- 8)E18 & $>$7E-3 & $<$500 & 1.1&1\\
$4A$&\object{HE0151--4326} & 2.775 &
2.4158 & -2 -- -1.5 & (3 -- 7)E16 & $>$7E-4 & $<$30 & $\sim$1.0&1 \\
$4B$&&&2.4158 & -3.2 -- -2.2 & (1 -- 10)E18 & $>$(0.1 -- 1)E-2 &
$<$300 & 1.4 -- 2.0 & 1\\
$4C$&&&2.4158 & -2 -- -1.5 & (1.5 -- 3.5)E16 & $>$7E-4 & $<$15 & $\sim$1.0 & 1\\
5&&&1.7315 & -2.3 & (5 -- 6)E18 & $>5$E-3 & $<$300 & 0.9 & 1\\
6&\object{HE0001--2340} & 2.26 &
1.6514 & -2.6 & (5 -- 6)E17 & $>$8E-3 & $<$20 & 0.8&1 \\
7&&&1.5770 & -2.0 -- -1.8 & (6 -- 8)E16 &$>$1E-3 & $<$20 & $\sim0.2$&1 \\
8&\object{HE0141--3932} & 1.80 &
1.7817 & -2.5 & (4 -- 5)E17 &$>$7E-3 &$<$20  & 1 &2\\
9&\object{HE2347--4342} & 2.88 &
2.8980 & $>$-1.0 & $\sim$1E17 & $>$1E-4 & $<$300 & 1 &3\\
10&&&1.7963 & -2.7 & $\sim$1E17 & $>2$E-2 & $<$4 & 0.6 &4\\
11&\object{Q0329--385} & 2.435 &
2.3520& $\sim$ -0.75 & $\sim$ 1E17 & $>$1E-4 & $<$300 & 2 -- 3&3 \\
\noalign{\smallskip}
\hline
\noalign{\smallskip}
\multicolumn{10}{l}{$^\dagger$Mean kinetic temperature in units of $10^4$ K.}\\
\multicolumn{10}{l}{$^a$References.~--- (1) present paper. 
(2) Reimers \etal\ 2005. (3) Levshakov \etal\ 2008. (4) Agafonova \etal\ 2007.}\\
\end{tabular}
\end{table*}

\section{Analysis of individual systems}
\label{sect-2A}

\subsection{Quasar \object{HE1347--2457} }
\label{sect-2-1}

\subsubsection{System at \zabs = 2.5745}
\label{sect-2-1-1}

The system at \zabs = 2.5745 consists of the neutral hydrogen Lyman series 
lines (L$_1$--L$_{10}$) 
and many lines of different metal ions (Fig.~\ref{fg_1}). 
Expected positions of the \ion{Fe}{ii} $\lambda\lambda$2382.76, 2600.17 lines 
coincide with strong telluric absorptions, 
\ion{Fe}{iii} $\lambda$1122.52 falls 
in a noisy part of the spectrum and cannot be extracted from the noise. 
The emission redshift of the quasar \object{HE1347--2457} is \zem = 2.578 
(based on \ion{C}{ii} emission), so that
the absorber is detached by only 300 \kms\ from the quasar. 
The profiles of the \ion{O}{vi} $\lambda\lambda1031, 1037$ absorption lines
differ from the profiles of other ionic 
transitions all of which exhibit similar line shapes.  
The observed intensities of low ionization (\ion{C}{ii}, \ion{Si}{ii}) 
and high ionization (\ion{C}{iv}, \ion{Si}{iv}) lines
do not differ from each other significantly implying
a rather low ionization parameter $U_0$, $\log(U_0) < -1.7$, for all
types of the incident ionizing spectra.  At such $U_0$, 
a considerable amount of \ion{O}{vi} can hardly
be produced.  Thus, this absorption system originates probably in a dense
and relatively cold gas 
(seen in lines of \ion{C}{ii}--\ion{C}{iv}, \ion{Si}{ii}--\ion{Si}{iv},
\ion{N}{iii}, \ion{Mg}{ii}, and \ion{Al}{iii})
embedded in a hot and highly ionized medium seen in \ion{O}{vi} and \ion{H}{i}
$\lambda\lambda$1215, 1025 lines.

\begin{table*}[t!]
\centering
\caption{Calculated column densities for atoms and ions identified
in metal-rich absorbers (for references, see Table~\ref{tbl-1}).
}
\label{tbl-2}
\begin{tabular}{lcccccc}
\hline
\hline
\noalign{\smallskip}
Redshift, & H\,{\sc i} & \hspace{-0.1cm}C\,{\sc ii} 
& N\,{\sc ii} & \hspace{-0.2cm}O\,{\sc i} & 
Mg\,{\sc ii} & \hspace{-0.1cm}Si\,{\sc ii} \\[-2pt]
\multicolumn{1}{c}{\zabs} &  & C\,{\sc iii} & \hspace{0.1cm}N\,{\sc iii} 
& O\,{\sc vi} & Fe\,{\sc ii} & Si\,{\sc iii} \\[-2pt]
&  & C\,{\sc iv} & N\,{\sc v} & & Al\,{\sc iii} & Si\,{\sc iv} \\
\noalign{\smallskip}
\hline 
\noalign{\medskip}
\multicolumn{7}{c}{\it \object{HE1347--2457} (\zem = 2.578)}\\
2.5745$^\ast$ &(9.5$\pm$0.3)E14 &(1.6$\pm$0.2)E13 & 
--- & --- & (1.3$\pm$0.2)E12 & (1.4$\pm$0.2)E12 \\[-2pt]
& &(1.9$\pm$0.3)E14 &(1.8$\pm$0.2)E13 & ---& ---&(4.7$\pm$0.4)E12  \\[-2pt]
& &(3.2$\pm$0.1)E13 &$<$5.0E11 & &(3.6$\pm$0.5)E11  & (3.7$\pm$0.3)E13 \\
1.7529$\ast$ &(1.0 -- 1.5)E16 &(1.5$\pm$0.1)E14 & 
--- & --- & (6.8$\pm$0.2)E12 &(9.2$\pm$1.0)E12 \\[-2pt]
 &  & ---& ---& ---& $<3.5$E11 &(5.5$\pm$0.5)E13 \\[-2pt]
 &  &(3.2$\pm$0.5)E14 & $<$1.0E13 & & (1.7$\pm$0.2)E12 & (4.5$\pm$0.5)E13 \\
1.5080$^\ast$ & (1.0 -- 1.5)E16 & ---&
 ---&  ---&
(1.1$\pm$0.1)E13 & (1.5$\pm$0.1)E13 \\[-2pt]
 & & --- & --- & --- & (2.1$\pm$0.2)E12 & ---\\[-2pt]
 & &(1.7$\pm$0.2)E14 & $<$7.0E12 & & (2.2$\pm$0.2)E12 &(5.4$\pm$0.5)E13 \\
\noalign{\smallskip}
\multicolumn{7}{c}{\it \object{HE0151--4326} (\zem = 2.775)}\\
{ 2.4158}$^\diamond$ &   \\[-2pt]
\multicolumn{1}{c}{ $A$}
&(2.2$\pm$0.1)E13 &$<5.0$E11& ---& ---& ---& --- \\[-2pt]
& &(1.5$\pm$0.2)E13& $<$1.1E13& --- & ---& $<$8.5E10  \\[-2pt]
& &(1.2$\pm$0.1)E13 & $<1.5$E12 & & --- & $<2.5$E11\\
\multicolumn{1}{c}{ $B$}
&(7.5$\pm$0.5)E15 & $<1.0$E12 & $<$4.0E12 & ---&  --- & { ---} \\[-2pt]
& &(6.5$\pm$1.0)E12 & $\leq$7.0E12 & --- & ---& $<$2.8E11  \\[-2pt]
& & $<$1.0E12 &  ---& & --- & $<2.5$E11  \\
\multicolumn{1}{c}{ $C$}
&(1.0$\pm$0.2)E13& ---& --- & ---& ---& --- \\[-2pt]
& &(6.5$\pm$0.5)E12 & $^\ddagger$(5 -- 6)E12 & $<4.5$E12& ---& ---\\[-2pt]
& &(7.0$\pm$0.5)E12 & --- & &--- & $<$2.5E11  \\
1.7315 & (0.9 -- 1.2)E16 &(1.0$\pm$0.1)E14 &
 ---& ---&(9.0$\pm$1.0)E12 & (1.0$\pm$0.1)E13 \\[-2pt]
& & ---& --- & --- & (3.4$\pm$0.3)E11 &(3.6$\pm$0.5)E13\\[-2pt]
& &(3.2$\pm$0.5)E14 &$<$1.5E13 & & (1.3$\pm$0.2)E12 & (3.8$\pm$0.5)E13\\
\noalign{\smallskip}
\multicolumn{7}{c}{\it \object{HE0001--2340} (\zem = 2.26)}\\
1.6514&(3.5 -- 4.5)E15&$^\ddagger$3.7E13&
 ---& $<$1.0E12 &(4.1$\pm$0.4)E12 & (1.8$\pm$0.2)E12 \\[-2pt]
& & --- & ---& --- & (2.0$\pm$0.2)E11 &{ (2.5$\pm$0.4)E12}\\[-2pt]
& &(4.7$\pm$0.2)E13 & --- & & $<$1.3E11 & (2.2$\pm$0.2)E12 \\
1.5770 & (0.9 -- 1.0)E14 &(7.6$\pm$0.7)E12 &
---& --- & $<$3.0E11 &  ---\\[-2pt] 
& & ---& --- & --- & ---& (5.4$\pm$1.2)E11 \\[-2pt]
&&(5.2$\pm$0.5)E13&$^\ddagger${(0.8 -- 1.1)E13}&&$<$1.3E11&(1.0$\pm$0.2)E12\\
\noalign{\smallskip}
\multicolumn{7}{c}{\it \object{HE0141--3932} (\zem = 1.80)}\\
1.7817 &(2.5 -- 3.0)E15&(1.64$\pm$0.08)E13&
 ---& --- &(1.2$\pm$0.2)E12 & (2.8$\pm$0.3)E12  \\[-2pt]
& & --- & --- & --- & (1.4$\pm$0.4)E11 & ---\\[-2pt]
& &(2.2$\pm$0.2)E13 & $<$1.4E12 & & (3.6$\pm$1.0)E11 & (4.5$\pm$0.4)E12 \\
\noalign{\smallskip}
\multicolumn{7}{c}{\it \object{HE2347--4342} (\zem = 2.88)}\\
2.8980$^\ast$ &(1.1 -- 2.3)E13 & $<$3.0E11& ---& ---& ---& ---\\[-2pt]
& &(5 -- 9)E12 & ---&(2.8$\pm$0.2)E14 & --- & --- \\[-2pt]
& &(1.2$\pm$0.1)E14 & (2.6$\pm$0.2)E13 & & --- & ---\\
1.7963 &(0.9 -- 1.2)E16 &$^\dagger$5.9E13 &
---& (1.0$\pm$0.2)E13 
&(0.9 -- 1.6)E13 & (1.4$\pm$0.1)E13 \\[-2pt] 
& & --- & ---& ---& (9.4$\pm$0.5)E11 &(4.9$\pm$1.0)E12\\[-2pt]
& &(1.2$\pm$0.1)E13 & --- & & (7.4$\pm$0.7)E11 &(2.9$\pm$0.3)E12 \\
\noalign{\smallskip}
\multicolumn{7}{c}{\it \object{Q0329--385} (\zem = 2.435)}\\
2.3520 &(0.7 -- 1.3)E13 &$<$5.0E11& ---& ---& ---& --- \\[-2pt]
& &$^\dagger$3.6E12 & $^\dagger$2.9E12&(1.2$\pm$0.1)E14 &--- & --- \\[-2pt]
& &(3.2$\pm$0.1)E13 &(4.0$\pm$0.2)E13 & &--- & --- \\
\noalign{\smallskip}
\hline
\noalign{\smallskip}
\multicolumn{7}{l}{$^\ast$Incomplete covering of the background light source.}\\
\multicolumn{7}{l}{$^\diamond$$A$: $-40<v<30$ \kms, 
$B$: $130 < v < 330$ \kms, $C$: $300 < v < 370$ \kms, see Fig.~\ref{fg_8}.}\\
\multicolumn{7}{l}{$^\ddagger$Deconvolved from Ly$-\alpha$ forest absorption.}\\
\multicolumn{7}{l}{$^\dagger$Calculated with the recovered
velocity and density distributions.}
\end{tabular}
\end{table*}

The absorption profiles of the
\ion{C}{iii} $\lambda$977 and \ion{H}{i} $\lambda1025$ 
lines\footnote{The \ion{H}{i} $\lambda$1215 line
is black because of blending with Ly-$\alpha$ from the
highly ionized system. 
An appropriate $N$(\ion{H}{i}) for this system is $\sim 10^{14}$ \cm,
i.e. its Ly-$\alpha$ is saturated but Ly-$\beta$ is not. }
have flat bottoms and reveal non-zero residual intensities 
at the line centers (0.02 and 0.01, respectively) 
pointing to the incomplete coverage of the background light source, 
i.e. the covering factor for 
\ion{C}{iii} is  ${\cal C}$(\ion{C}{iii})$ = 0.98$ and 
for hydrogen ${\cal C}$(\ion{H}{i})$ = 0.99$. 
Due to the detection of many lines of neutral hydrogen its
column density can be estimated quite accurately: 
$N$(\ion{H}{i}) $= (9.5\pm0.3)\times10^{14}$ \cm. 
A clear (unblended) doublet of \ion{C}{iv} $\lambda\lambda$1548, 1550 
allows us to estimate ${\cal C}$(\ion{C}{iv})$ = 0.99$ and
$N$(\ion{C}{iv}) $= (3.2\pm0.1)\times10^{13}$ \cm. 
Because of the line blending, low S/N and/or the presence of
a single unsaturated line only (like \ion{Si}{iii} $\lambda1206$ and 
\ion{N}{iii} $\lambda989$) accurate covering factors 
for other ions cannot be determined. 
Their limiting values are as follows:
$0.98 > {\cal C}$(\ion{C}{ii}/\ion{Si}{ii}/\ion{Mg}{ii}) $> 0.6$, 
$0.98 > {\cal C}$(\ion{Si}{iii}/\ion{N}{iii}/\ion{Al}{iii}) $> 0.7$, and
$0.98 > {\cal C}$(\ion{Si}{iv}) $> 0.8$.
We note that covering factors may be slightly different for each of the two
components comprising the absorption system under consideration. 

With such uncertainties in covering factors the shape 
of the ionizing spectrum cannot be restored uniquely. 
Nevertheless, some conclusions about the SED can be obtained.
The incomplete covering of the background light source indicates that the system
is located close to the quasar. Common first guess for the SED of the
outcoming quasar radiation is a power law, $\nu^{-\alpha}$. 
However, pure power law spectra 
are not consistent with the observed line intensities. Namely, for all
tried spectral indices $\alpha$, $-2.0 < \alpha < -0.5$,
the mean ionization parameter $U_0$ is low and the corresponding
fraction of \ion{C}{iv} is low as well, 
which makes the abundance of carbon,
$$
{\rm C} \sim \frac{\rm C\,\scriptstyle IV}{\rm H\,\scriptstyle I}
\frac{\Upsilon_{\scriptstyle \rm H\,\scriptscriptstyle I}}
{\Upsilon_{\scriptstyle \rm C\,\scriptscriptstyle IV}}\, ,
$$
to be very high, $4-5$ times solar value 
(and the abundance of magnesium even higher). 
With such a high metal content it is impossible to describe the
observed profiles of hydrogen lines: in particular the central parts of the
synthetic profiles of the $\lambda$972.53 and $\lambda$937.80 
lines look too wavy following the pattern exhibited by metal ions.
In order to decrease the metallicity, the ratio 
$\Upsilon_{\scriptstyle \rm H\, \scriptscriptstyle I}/
\Upsilon_{\scriptstyle \rm C\, \scriptscriptstyle IV}$ 
should be lowered, and this occurs when
the observed profiles of \ion{C}{ii}/\ion{C}{iv} and 
\ion{Si}{ii}/\ion{Si}{iii}/\ion{Si}{iv} lines 
are described with higher values of $U_0$.  
As it was the case 
with the associated systems from Paper I, to fit
to these conditions the ionizing spectra should have a break at 4 Ryd.  
The depth of the break depends on the adopted covering factors, 
but the intensity decrease of at least one order of magnitude at $E > 4$ Ryd 
is required. 
On the other hand, a too deep drop of the intensity 
(the depth of the break $> 1.3$ dex)
leads to the overproduction of \ion{Si}{iii} compared 
to the observed profile even 
at the lower limit of ${\cal C}$(\ion{Si}{iii}). 
A small difference between the \ion{C}{iv}/\ion{C}{ii}  and 
\ion{Si}{iv}/\ion{Si}{ii} ratios makes
preferable the power indices between 1 and 4 Ryd of 
$\alpha \sim  1.2-1.4$.

\begin{table*}[t!]
\centering
\caption{Abundance patters for the absorbers listed in Table~\ref{tbl-1}.
[X/Y] means  $\log (N_{\rm X}/N_{\rm Y}) - \log (N_{\rm X}/N_{\rm Y})_\odot$.
Point estimates have uncertainties of 0.05 dex.
}
\label{tbl-3}
\begin{tabular}{ccccccccc}
\hline
\hline
\noalign{\smallskip}
\# & \multicolumn{1}{c}{\zabs} & [C/H] & [N/C] & [O/C] & [Mg/C]  
& [Al/C] & [Si/C] & [Fe/C] \\
\noalign{\smallskip}
\hline 
\noalign{\medskip}
1&2.5745 & 0.4 -- 0.5 & -0.2 & --- &  0 -- 0.1 & $0.0$ & -0.3 & ---\\ 
2&1.7529 & 0.1 -- 0.2 &$<0.2$ & ---& -0.1 -- 0.1 & 0.0 & -0.3&$<$ -0.25\\
3&1.5080 & 0.1 -- 0.2 &$<$0.3 &---& $\ga$0.2 & $\ga$0.15 & 0.0&$\ga$0.3 \\
$4A$&2.4158 & 0.2 -- 0.5 & $\la$0.5 &---&---&---& $<$ -0.6 &--- \\
$4B$&2.4158 & -1.5 -- -2.5 & $\la$0.5& ---&---&---&$<$ -0.5 & --- \\
$4C$&2.4158 & 0.2 -- 0.3 & $\la$0.5 &---&---&---& $<$ -0.6 &--- \\
5&1.7315 & 0.2 -- 0.35 & $<$0.1 &---& 0.25 & 0.0 & -0.25 & -0.2\\
6&1.6514 &0.40 -- 0.45 &---&$<$0.1&0.1 -- 0.15&$<-$0.6& -0.8&$\la$ -0.3\\
7&1.5770 & $\sim1.0$ & 0.3 -- 0.5 & --- & $<$ 0 & $<$ 0 & -1.0 -- -0.7 & --- \\
8&1.7817 & 0.15 & $<$0.0 & --- & 0.15 & 0.15 & -0.05 & 0.15 -- 0.25\\
9&2.8980  & $\sim1.0$ & -0.3 & 0.0 -- 0.5 & ---&---&---&--- \\
10&1.7963 & 0.5 & --- & 0.0 & 0.2 -- 0.4 & 0.12 & 0.0 & -0.55 \\
11&2.3520  & 0.3 -- 0.5 & 0.4 & 0.0&---&---&---&--- \\
\noalign{\smallskip}
\hline
\end{tabular}
\end{table*}

As already noted, at high metallicities (above solar) the relative
element contents, especially those 
of the most abundant elements carbon and oxygen,
affect quite strongly the calculated ion fractions (upon which the procedure of
spectral shape estimation is based). 
Lines of different oxygen ions are not available
in the \zabs = 2.5745 system, and the unknown oxygen 
content gives an additional degree of freedom in the procedure of the 
spectral shape estimation.  These considerations restrict
the acceptable spectral shapes as shown by
the shaded area in Fig.~\ref{fg_2}. 
The values of the physical parameters obtained with the MCI routine
are given in Tables~\ref{tbl-1}~-- \ref{tbl-3}. 
The column densities in Table~\ref{tbl-2} are calculated 
with the upper limits of the covering factors, i.e. real column 
densities may be up to 50\% higher if, for instance, ${\cal C} \simeq 0.6$. 
The gas density of $5\times10^{-3}$ \cmm, and  the linear size
of 25 pc given in Table~\ref{tbl-1} are estimated with the mean 
intergalactic value of $J_{912}$. 
However, for an absorber located close to the quasar 
the intensity of the ionizing continuum $J_{912}$ will be
enhanced making the corresponding gas density
higher and the linear size smaller. 
For example, $J_{912}$ in the vicinity of
the associated system at \zabs = 2.1470 towards 
\object{HE0141--3932}  (Sect. 2.2.1 in Paper~I) 
was enhanced by two orders of magnitude as compared to the 
mean intergalactic level. With such an
intensity the linear size of the present absorber  
would be of a sub-parsec scale. 

In spite of all uncertainties inherent to the present system, the relative
abundance ratios retain a quite stable pattern 
(Table~\ref{tbl-3}).
Namely, the abundances of carbon, magnesium and aluminium 
are strongly oversolar reaching
$2.5-3$ solar values whereas silicon and nitrogen are
considerably underabundant compared to carbon
(by $0.2-0.4$ dex depending on the shape 
of the ionizing spectrum and the
adopted covering factors for silicon and nitrogen ions). 

The derived underabundance of nitrogen to carbon, [N/C] $< -0.2$, at highly
oversolar metallicity 
does not comply with models of chemical evolution of gas in the
central parts of QSO-hosting galaxies (Hamann \& Ferland 1999) and models
of galactic chemical enrichment (Calura \& Matteucci 2004)
all of which predict significant overabundance of nitrogen relative to other
elements at metallicities above solar. However, nitrogen measurements in
planetary nebulae show that even at high metallicities nitrogen can be both
under- and overabundant 
(Aller \& Czyzak 1983; Perinotto 1991; Richer \& McCall 2008). 
Low ratios [N/C] $< 0$ at near-solar  metallicities were reported also for
intergalactic absorbers by
Reimers \etal\ (2005) and  Jenkins \etal\ (2005).

Quite peculiar in the abundance pattern of the system under study
is its low content of silicon. In optically-thin quasar absorbers
the standard finding is
[Si/C] $ > 0$ which is attributed
to the SNe Type~II stars as the main source of silicon enrichment
(Songaila \& Cowie 1996; Simcoe \etal\ 2006).
In the \zabs = 2.5745 system the contents of 
C, Mg and Al follow the solar pattern, 
thus we can conclude that it is not carbon which is enhanced but silicon 
which is in fact deficient.
Taking into account high metallicity of the absorbing gas, 
it is conceivable to assume that this deficit is caused by depletion of 
silicon into dust. 
It is well known that convective envelopes of post-main sequence 
stars (Red Giants, Red Supergiants and Asymptotic Giant Branch stars) 
as well as planetary nebulae forming around these stars are
places of the effective dust condensation. 
In the oxygen-rich environment, C/O $< 1$,
the main dust species are iron and silicates consisting of olivine
Mg$_{2x}$Fe$_{2(1-x)}$SiO$_4$, pyroxen
Mg$_x$Fe$_{1-x}$SiO$_3$ and quartz SiO$_2$ 
(Gail \& Sedlmayr 1999; Jeong \etal\ 2003; Ferrarotti \& Gail 2006).
Proportions of olivine, pyroxen and quartz among the silicate
dust grains are unknown, thus magnesium could also be depleted 
along with silicon and oxygen.
However, there are reasons to suppose
that depletion of magnesium~--- if any~--- is in fact much 
less pronounced than that of silicon (see Sect.~\ref{sect-3-2}).  

\begin{figure*}[t]
\vspace{0.0cm}
\hspace{-0.2cm}\psfig{figure=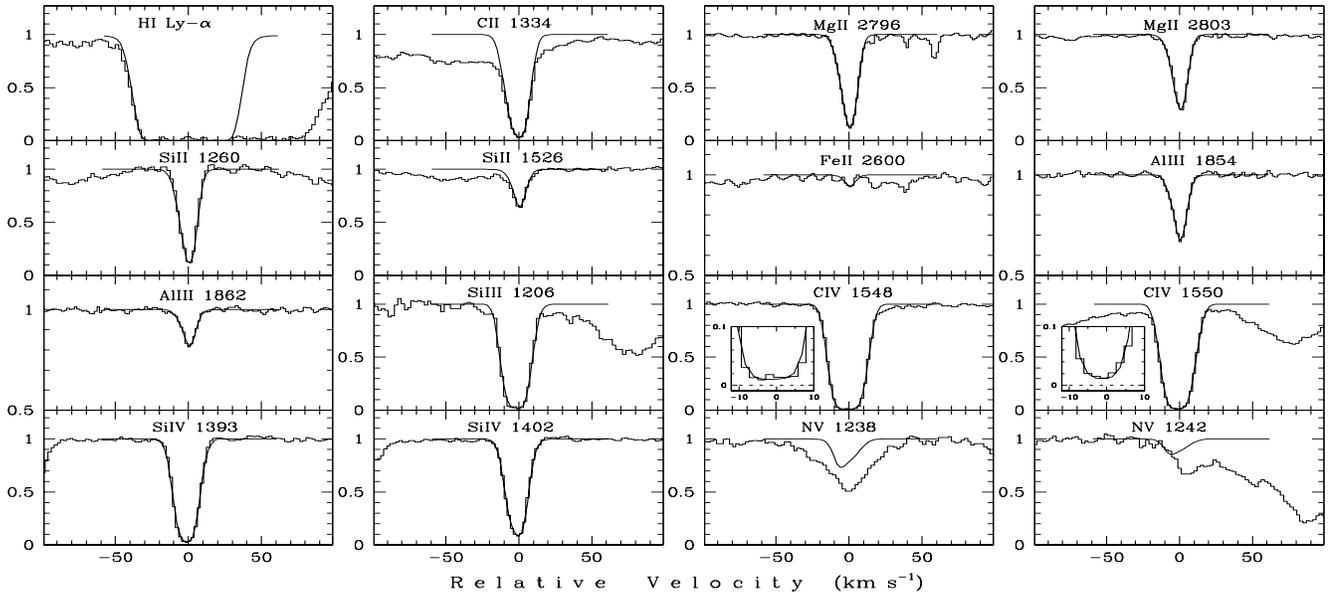,height=14cm,width=18cm}
\vspace{-5.8cm}
\caption[]{Same as Fig.~\ref{fg_1} but for
the \zabs = 1.7529 system (\object{HE1347--2457}).
The vertical axis is normalized intensities.
The zero radial velocity is fixed at $z = 1.7529$.
}
\label{fg_3}
\end{figure*}

Taking into account high metallicity, 
relative element ratios, small linear size and proximity to the quasar, 
we can suppose that the absorber is formed by fragment(s) of planetary 
nebula(e) or AGB-star envelope(s) ejected into the intergalactic space 
by the quasar wind. Since gas and dust particles
have different velocities in the stream, dust-gas separation occurs, and
the gas carried away into the halo of the quasar host galaxy
or even into the intergalactic   space becomes 
dust-free. However, elements condensed into the dust remain deficient
even in the dust-free absorber.

\subsubsection{System at \zabs = 1.7529}
\label{sect-2-1-2}

The absorber at \zabs = 1.7529 exhibits a
saturated \ion{H}{i} $\lambda1215$ line and strong lines of 
many carbon and silicon ions, all having very simple profile shapes
(Fig.~\ref{fg_3}). The blue and red wings of the hydrogen line look different,
and the metal lines are shifted relative to its center. 
Additionally, the flat bottom
lines of the \ion{C}{iv} $\lambda\lambda1548, 1550$ doublet and 
\ion{Si}{iii} $\lambda1206$ show non-zero residual intensity at the line
centers which means that the background radiation is leaking. 
The covering factor for these lines can be easily set as 
${\cal C}$(\ion{C}{iv}) = 0.99, ${\cal C}$(\ion{Si}{iii}) = 0.98.
Clear profiles are also available for the doublets
\ion{Si}{iv} $\lambda\lambda1393, 1402$ and 
\ion{Mg}{ii} $\lambda\lambda2796, 2803$, so it is possible to estimate the
accurate covering factors for these lines as well: 
${\cal C}$(\ion{Si}{iv}) = 0.98, and
${\cal C}$(\ion{Mg}{ii}) = 0.98.
Assuming that ions \ion{C}{ii} and \ion{Si}{ii} trace the same gas as 
\ion{Mg}{ii}, we  set
${\cal C}$(\ion{C}{ii}) = ${\cal C}$(\ion{Si}{ii}) = 0.98.

The incomplete covering arises probably due to  
micro-lensing of the background quasar by some intervening galaxy.
This naturally supposes a very small size of the absorber. 
In spite of the presence of only one saturated line of neutral hydrogen,
it is possible to estimate its column density with a
relatively high accuracy (up to 30\%) since the velocity
dispersion of the gas can be well
restricted by the numerous metal lines.

Given the ions of different ionization stages of the same element 
(\ion{C}{ii}, \ion{C}{iv}, \ion{Si}{ii}, \ion{Si}{iii}, \ion{Si}{iv}),
the spectral shape of the ionizing background radiation can be restored
rather accurately (see Fig.~\ref{fg_4}). 
In order to describe the observed intensities, $I_\lambda$, of all ions 
at the same value of $U_0$,  
the ionizing spectrum should have a soft slope with the index 
$\alpha = 1.7-1.8$
between 1 and 4 Ryd, and a drop in the intensity at 4 Ryd by 
$3-4$ times. 

The expected position of the neutral oxygen line 
\ion{O}{i} $\lambda1302.16$ is blended with a strong forest
absorption which prevents setting a limit on oxygen abundance. 
This unknown abundance represents
an additional degree of freedom in the procedure 
of the spectral shape recovering: with scaled
down (relative to carbon) oxygen abundance the above conditions are
fulfilled for the spectral shapes which are softer at $E > 4$ Ryd. 
An acceptable range for the ionizing spectra
is shown in Fig.~\ref{fg_4}.
In all cases the carbon content comes slightly
oversolar ($1.1-1.5$ solar values). 
The physical parameters of the absorbing gas are given in 
Table~\ref{tbl-1}, column densities
and the element abundances~--- in Tables~\ref{tbl-2} and \ref{tbl-3},
respectively. 

The abundance pattern is almost identical to that found in the previous 
\zabs = 2.5745 system: silicon
is significantly depleted relative to carbon, magnesium and aluminium.
The absolute abundance of magnesium depends on the depth of the break at
4 Ryd: more softer spectra give lower Mg content,
so that the ratio [Mg/C] can be by $\sim 0.1$ dex both below and above zero.
A clear continuum window at the expected position of the
\ion{Fe}{ii} $\lambda2600$ line allows us to set an upper limit
on the iron content: [Fe/C] $< -0.25$, i.e. iron
is clearly underabundant relative to carbon.
Unfortunately, both nitrogen lines 
\ion{N}{v} $\lambda\lambda1238.82, 1242.80$ are
contaminated by forest absorption, and only a not very instructive 
upper limit on the nitrogen content can be set, [N/C] $< 0.2$.

\begin{figure}[t]
\vspace{-1.0cm}
\hspace{-1.5cm}\psfig{figure=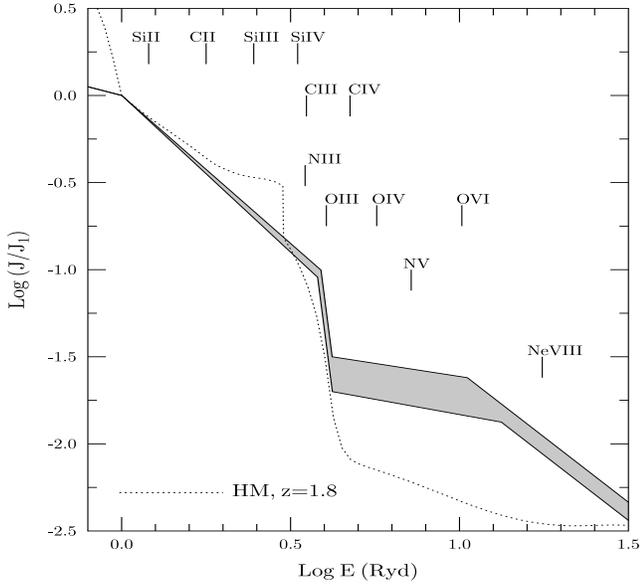,height=13cm,width=12cm}
\vspace{-4.5cm}
\caption[]{Same as Fig.~\ref{fg_2} but for
the \zabs = 1.7529 system (\object{HE1347--2457}). 
}
\label{fg_4}
\end{figure}

The high metallicity and the relative ratios of C, Si, Mg, Al, 
with silicon and probably
magnesium depleted compared to carbon and aluminium, suggest again 
(as it was for the previous \zabs = 2.5745 absorber) 
the dust-forming envelopes of the post-main sequence stars or planetary nebulae
as a possible source of the observed absorption. 
The underabundance of iron relative to carbon can 
be explained by different reasons: it may be intrinsic due to a
relatively low contribution from SNe~Ia to the enrichment of gas, 
or it may be produced in the AGB-stars themselves through the depletion 
onto dust grains and/or $s$-process (Herwig 2005).

Clear and strong lines of \ion{Mg}{ii} make it possible
to conduct one more test whether the AGB-stars are at play or not.
Theory predicts that in
the AGB-stars with masses $M > 4 M_\odot$ 
the heavy magnesium isotopes
$^{25}$Mg and $^{26}$Mg should be enhanced 
(Karakas \& Lattanzio 2003).
Solar isotopic composition is $^{24}$Mg:$^{25}$Mg:$^{26}$Mg =
78.99:10.00:11.01, but observations
of red giant stars in globular clusters show in some cases 
much higher input of $^{25}$Mg and $^{26}$Mg
(Yong \etal\ 2003, 2006). 
Due to secondary character of $^{25}$Mg and $^{26}$Mg 
the content of these isotopes is expected to scale with metallicity.

The wavelengths of the resonance transitions of \ion{Mg}{ii} are
2796.3553 \AA\ for $^{24}$Mg, 2796.3511 \AA\ for $^{25}$Mg,
and 2796.3473 \AA\ for $^{26}$Mg (Morton 2003).
For solar isotopic composition 
Morton gives the weighted mean value 
$\lambda_{\rm rest} = 2796.3543$ \AA. 
If the abundances of the heavy
isotopes are enhanced, then the wavelength $\lambda_{\rm rest}$
becomes smaller, i.e.
the observed at a given redshift \ion{Mg}{ii} lines are shifted toward lower 
radial velocities as compared with their positions expected for
the solar isotopic composition ratio. 

\begin{figure}[t]
\vspace{-1.0cm}
\hspace{-0.2cm}\psfig{figure=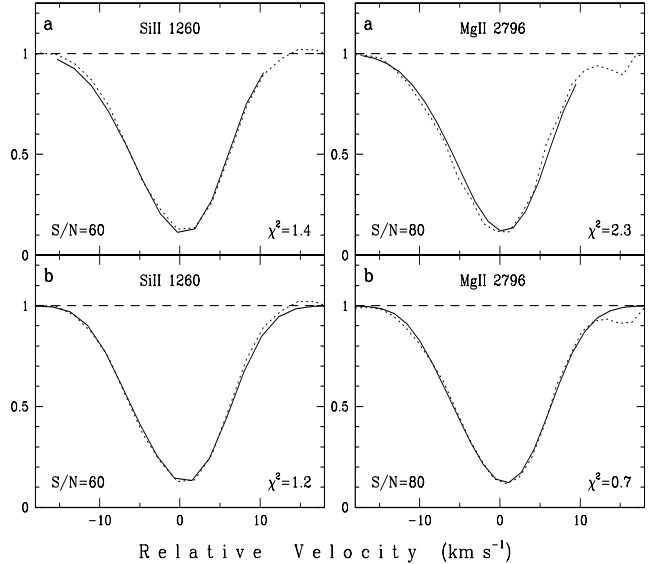,height=11cm,width=9cm}
\vspace{-2.5cm}
\caption[]{\ion{Mg}{ii} in
the \zabs = 1.7529 system (\object{HE1347--2457}). 
Panel {\bf a}: observed (points) and synthetic (solid lines)
profiles when \ion{Mg}{ii} lines are centered using 
the solar composition of the isotopes $^{24}$Mg:$^{25}$Mg:$^{26}$Mg = 
79:10:11. 
Panel {\bf b}: observed (points) and synthetic (solid lines)
profiles when \ion{Mg}{ii} lines are shifted by 0.6 \kms\ 
which corresponds to the inverse isotope composition
$^{24}$Mg:$^{25}$Mg:$^{26}$Mg = 11:10:79.
The corresponding signal-to-noise (S/N) mean value per resolution
element at the continuum level and the obtained $\chi^2_{\rm min}$ per
degree of freedom are shown.  
}
\label{fg_5}
\end{figure}

In our approach to calculate synthetic profiles we assume
that all lines are produced by the same absorbing gas with fluctuating
density and velocity fields. Profiles of specific ions may look 
different due to different responses of ions (ion fractions)  
to the local gas density which in the optically
thin case are  determined entirely by the spectral energy distribution
of the ionizing background. This means that profiles~--- in spite of being
different~--- should retain definite consistency governed 
by both the distributions of gas
density and velocity along the line-of-sight and by the ionizing background.
However, with magnesium lines centered using the solar composition wavelengths
the synthetic profiles of \ion{Mg}{ii} and other low-ionization transitions
\ion{C}{ii} $\lambda1334$ and \ion{Si}{ii} $\lambda1260$
look a bit incoherent (Fig.~\ref{fg_5}{\bf a}).

\begin{figure*}[t]
\vspace{0.0cm}
\hspace{-0.2cm}\psfig{figure=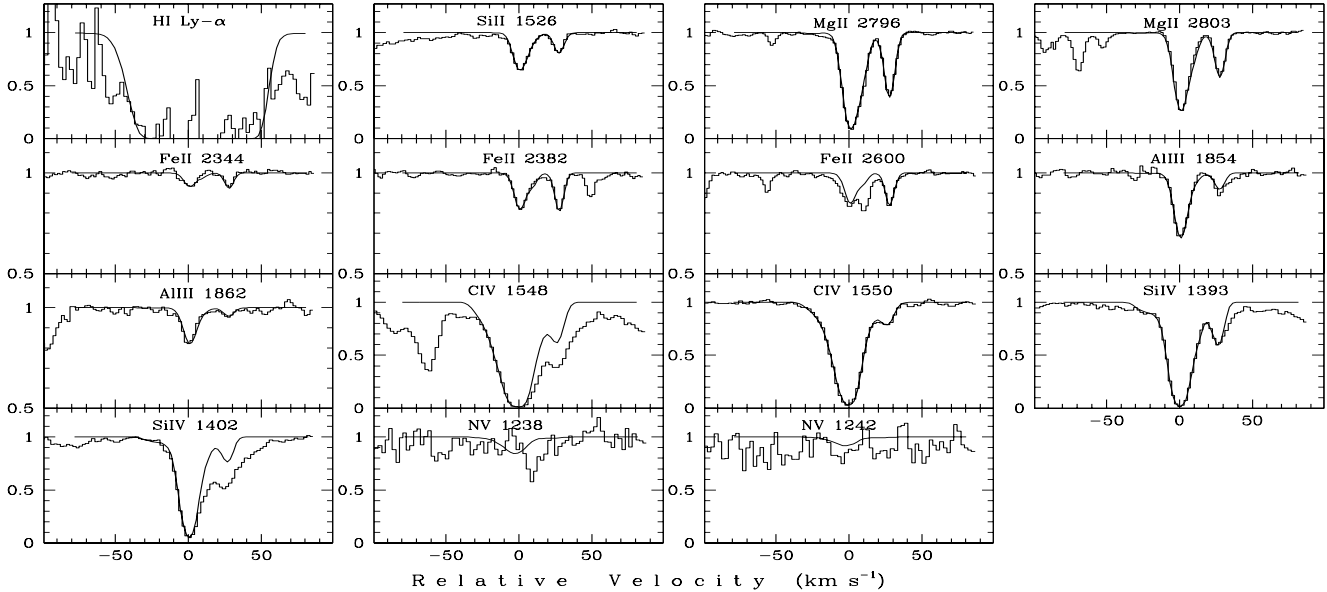,height=14cm,width=18cm}
\vspace{-6.0cm}
\caption[]{Same as Fig.~\ref{fg_1} but for
the \zabs = 1.5080 system (\object{HE1347--2457}).
}
\label{fg_6}
\end{figure*}

On the contrary, when the observed profiles of
\ion{Mg}{ii} are shifted by, e.g., 0.6 \kms\ towards higher radial
velocities 
(which would correspond to the inversion of the solar ratio 
of $^{24}$Mg to $^{26}$Mg, i.e. 
$^{24}$Mg:$^{25}$Mg:$^{26}$Mg = 11:10:79), the quality
of the fitting improves considerably (Fig.~\ref{fg_5}{\bf b}). 

However, with the available
spectral data this test is to a certain extent qualitative: 
positive shift of magnesium lines
is unambiguously preferred, but the accurate value of this 
shift cannot be determined since
the fittings with \ion{Mg}{ii} lines shifted by 
0.5 \kms\ and by 0.7 \kms are statistically indistinguishable.
In fact, accurate measurements of
the isotopic composition by means of a line
shift require a specially observed and processed quasar spectrum
(e.g., Levshakov \etal\ 2007; Molaro \etal\ 2008) and
cannot be carried out with the present spectrum of \object{HE1347--2457}.

As already mentioned above, the linear size of the
\zabs = 1.7529 system is probably very small.
The upper limit (assuming the absorber is intergalactic) is $L < 500$ pc.
The continuum window at the expected
position of the \ion{C}{ii}$^\ast$ $\lambda1335$ line allows us
to set an upper limit on its column density,
$N$(\ion{C}{ii}$^\ast$) $\la 9.0\times10^{11}$ \cm\
which results in an upper limit
on the gas density of $n_{\rm \scriptscriptstyle H} \la 0.14$ \cmm
(see Sect. 2.2.1 in Paper~I for details of calculations). 
With the total hydrogen column density of $(5-10)\times10^{18}$ \cm\
this gives $L \ga 12$ pc.

\subsubsection{System at \zabs = 1.5080}
\label{sect-2-1-3}

This system exhibits one saturated and very noisy 
hydrogen line (lies at the edge of observational wavelength range) 
accompanied by strong lines of metal ions in different ionization
stages (Fig.~\ref{fg_6}). 
Clear central parts of the doublets \ion{C}{iv} $\lambda1548, 1550$
and \ion{Si}{iv} $\lambda1393, 1402$ 
allow us to calculate accurately the corresponding covering
factors: ${\cal C}$(\ion{C}{iv}) = ${\cal C}$(\ion{Si}{iv}) = 0.99. 
The same covering factors were adopted for all other lines.
Given only one pair of the subsequent ions, \ion{Si}{ii} and 
\ion{Si}{iv} 
(\ion{C}{ii} $\lambda1334$ is 
blended with a deep forest absorption and cannot be deconvolved)
and unknown relative element abundances, it is impossible to conclude about the
SED of the ionizing radiation. 
However, 
the strong \ion{C}{iv} and the pronounced \ion{Fe}{ii} lines 
point to a spectrum at least as hard at $E > 4$ Ryd as
that estimated from the previous \zabs = 1.7529 system 
(Fig.~\ref{fg_4})~--- otherwise an overabundance
of iron to carbon significantly exceeds solar value, [Fe/C] $\ga 1.0$. 

For the same reason spectra with the indices 
$\alpha \sim 1.5-1.8$ between 1 and 4 Ryd are preferable.
With such spectra, abundances of carbon and silicon come out to be slightly
oversolar, however without deficit of silicon 
(cf. with column densities in the \zabs = 1.7529
system): [C/H] $\sim$ [Si/H] $\sim 0.1-0.2$, 
whereas magnesium, aluminium and iron are strongly 
enhanced relative to carbon: [Mg/C] $> 0.2$, 
[Al/C] $> 0.15$, [Fe/C] $> 0.3$, with very probable 
overabundance of iron to magnesium
as well, [Fe/Mg] $> 0.1$. 
There is a continuum window at the expected position 
of the \ion{N}{v} $\lambda1238$ line
which provides a conservative upper limit
on the nitrogen abundance [N/C] $< 0.3$. 
The physical parameters, column densities and derived abundances are given in
Tables~\ref{tbl-1}, \ref{tbl-2} and \ref{tbl-3}.
Very similar abundances were previously obtained for the narrow-line
associated system at \zabs = 1.7817
towards \object{HE0141--3932} (\#~8 in
Tables~\ref{tbl-2} and \ref{tbl-3}, 
see also Sect. 4.2 in Reimers \etal\ 2005). 
Such pattern~---
with iron enhanced relative  to $\alpha$-elements~--- 
is often observed in giants in local dwarf galaxies 
(Venn \etal\ 2004; Bonifacio \etal\ 2004; 
Tautvai\v{s}ien\.e \etal\ 2007)
and was also registered in some metal-rich giant stars in the Milky Way 
(Fabbian \etal\ 2005). 

As it was the case for the \zabs = 1.7529 system (Sect. \ref{sect-2-1-2}), 
simultaneous fitting of all lines
in the present system results in systematically shifted low-ionization lines
when \ion{Mg}{ii} is centered using the solar isotopic ratio
(Fig.~\ref{fg_7}{\bf a}), whereas
\ion{Mg}{ii} lines shifted by $\sim$0.6 \kms\ redward produce
almost perfect fitting (Fig.~\ref{fg_7}{\bf b}).
Thus, the enhanced content of heavy Mg isotopes 
seems very probable which means that AGB-stars 
contributed to the metal enrichment of gas in the \zabs = 1.5080 system.

\subsection{Quasar \object{HE0151--4326} }
\label{sect-2-2}

\subsubsection{System at \zabs = 2.4158}
\label{sect-2-2-1}

The absorbing complex at \zabs = 2.4158 was firstly mentioned in
Aracil \etal\ (2004). It
is spread over 350 \kms\ and can be divided into three parts 
(Fig.~\ref{fg_8}): two systems with weak \ion{H}{i} Ly-$\alpha$ lines
and pronounced metal lines of \ion{C}{iii} $\lambda977$ 
and \ion{C}{iv} $\lambda\lambda1548, 1550$ 
centered at $v = 0$ \kms\ (subsystem $A$) and $v = 347$ \kms\
(subsystem $C$, \zabs = 2.4196), and a subsystem $B$ comprising a 
strong \ion{H}{i} absorption in the range $140 < v < 340$ \kms\
(\zabs = 2.4180) accompanied  by only one prominent metal line~--  
\ion{C}{iii} $\lambda977$. 
That the observed absorptions are due to \ion{C}{iii} is quite certain: 
lines are narrow (i.e. cannot be hydrogen lines from the
Ly-$\alpha$ forest), aligned in velocity with the hydrogen absorption, 
and there are no any plausible metal contaminants from the 
identified intervening systems.

In the subsystems $B$ and $C$, there are weak and narrow lines
at the expected positions of the  \ion{N}{iii} $\lambda989$  (in $C$~--
blended with broad and shallow \ion{H}{i} forest absorption). 
No candidate for blending from other systems
was found, so quite probably these weak absorptions are indeed due to
\ion{N}{iii} $\lambda989$. 
From the  \ion{N}{v} doublet,  only a small unblended part of 
the \ion{N}{v} $\lambda1238$ line is present in the subsystem $A$. 
The \ion{O}{vi} doublet has an unblended part in the
\ion{O}{vi} $\lambda1037$ line seen in the subsystem $C$. 

The \ion{C}{ii} $\lambda1334$ lines coincide with a broad and shallow
intrinsic quasar absorption of \ion{H}{i} Ly-$\alpha$
(\object{HE0151--4326} is a mini-BAL quasar), 
but the absence of any pronounced
features at the expected positions of the \ion{C}{ii} 
lines in all three subsystems leads to the conclusion that these lines
should be very weak. 
Continuum windows at the positions of the silicon lines 
\ion{Si}{iii} $\lambda1206$ and \ion{Si}{iv} $\lambda1393$ 
give upper limits on their column densities. 
Unfortunately, both lines of \ion{Mg}{ii} $\lambda2796, 2803$ 
coincide with strong telluric absorptions. 

With ions available in all three sybsystems it is not
possible to restore the spectral shape of
the underlying ionizing background uniquely:
the observed line profiles can be described with UVB spectra ranging from pure
power laws to power laws with intensity break of different depths 
at 4 Ryd.
The blue wing of the \ion{H}{i} Ly-$\alpha$ line in the subsystem $A$ 
turned out to be inconsistent with the
assumption of constant metallicity throughout the absorber. 
The blue wing of the synthetic
Ly-$\alpha$ profile was then calculated on base of density 
and velocity distributions
restored from metal ions and the red part of the \ion{H}{i} Ly-$\alpha$ 
assuming a constant metal content of the absorbing gas. 

The column densities of all ions are listed in Table~\ref{tbl-2}.
The subsystems $A$ and $C$ reveal almost identical ratios 
\ion{C}{iv}/\ion{C}{iii} $\sim 1$ and 
\ion{C}{iv}/\ion{H}{i} $\sim 0.5$
indicating that these absorbers have very similar 
ionization states and metallicities.
In contrast, the absorber $B$ with its 
ratio \ion{C}{iv}/\ion{C}{iii} $< 0.15$ and
more than two orders 
of magnitude higher column density of neutral hydrogen,
$N$(\ion{H}{i}) = $7.5\times10^{15}$ \cm,
seems to be different both in physical state of the gas, 
and in metal content. 

In spite of the uncertainty in the UVB shape, element abundances retain  
a remarkably stable pattern (Table~\ref{tbl-3}).  
In particular, in the subsystems $A$ and $C$
at the value of the ionization parameter $U$ corresponding to 
$\Upsilon_{\scriptstyle \rm C\,\scriptscriptstyle IV}/
\Upsilon_{\scriptstyle \rm C\,\scriptscriptstyle III} \sim 1$
($\Upsilon_i$ is a fraction of ion $i$) all ionizing spectra
give oversolar abundance of carbon which can reach a few
solar values (the harder at $E > 4$ Ryd spectrum the higher carbon
content). 
On the other hand, the fractions of silicon ions \ion{Si}{iii} and 
\ion{Si}{iv} at $U$ corresponding to 
$\Upsilon_{\scriptstyle \rm C\,\scriptscriptstyle IV}/
\Upsilon_{\scriptstyle \rm C\,\scriptscriptstyle III} \sim 1$
are such high that only a significantly underabundant
silicon can comply with upper limits on 
$N$(\ion{Si}{iii}) and $N$(\ion{Si}{iv}): [Si/C] $< -0.6$. At this $U$
for all spectra we also have
$\Upsilon_{\scriptstyle \rm C\,\scriptscriptstyle III} =
\Upsilon_{\scriptstyle \rm N\,\scriptscriptstyle III}$.
If  the line in the subsystem $C$ is indeed due to
\ion{N}{iii} $\lambda989$ absorption, then
we obtain a clear overabundance
of nitrogen to carbon, [N/C] $\sim 0.5$.  

\begin{figure}[t]
\vspace{-1.0cm}
\hspace{-0.2cm}\psfig{figure=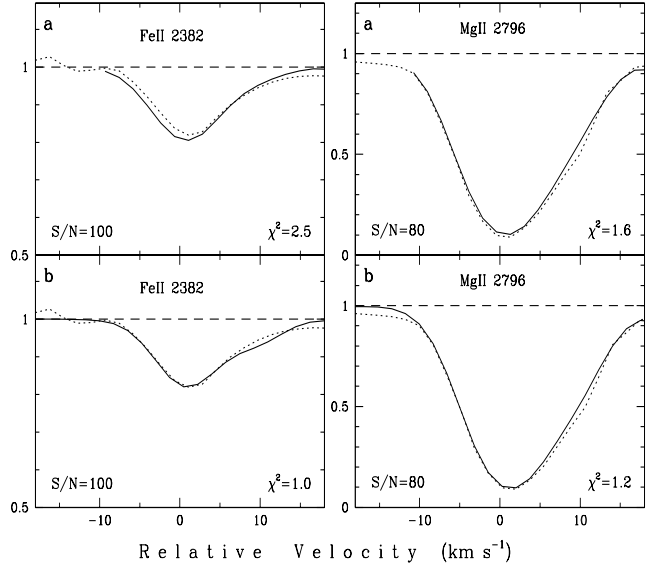,height=11cm,width=9cm}
\vspace{-2.5cm}
\caption[]{Same as  Fig.~\ref{fg_5} but for
the \zabs = 1.5080 system (\object{HE1347--2457}).
In contrast to Fig.~\ref{fg_5}{\bf a} where both lines \ion{Si}{ii} and
\ion{Mg}{ii} are equivalently strong, here \ion{Fe}{ii} is significantly
weaker than \ion{Mg}{ii} and, hence, the weight of \ion{Mg}{ii} dominates
in the minimization of the objective function leading to a prominent
shift of the \ion{Fe}{ii} line.
}
\label{fg_7}
\end{figure}

Unfortunately, the upper limit on $N$(\ion{O}{vi})  from  the subsystem $C$
cannot be translated into a reasonable bound on the oxygen abundance:
the \ion{O}{vi} fraction is extremely sensitive 
to the energy level above 4 Ryd, but
since this level cannot be estimated from the available ions, variations
of the oxygen abundance for applicable ionizing spectra 
exceed an order of magnitude.

\begin{figure*}[t]
\vspace{0.0cm}
\hspace{-0.2cm}\psfig{figure=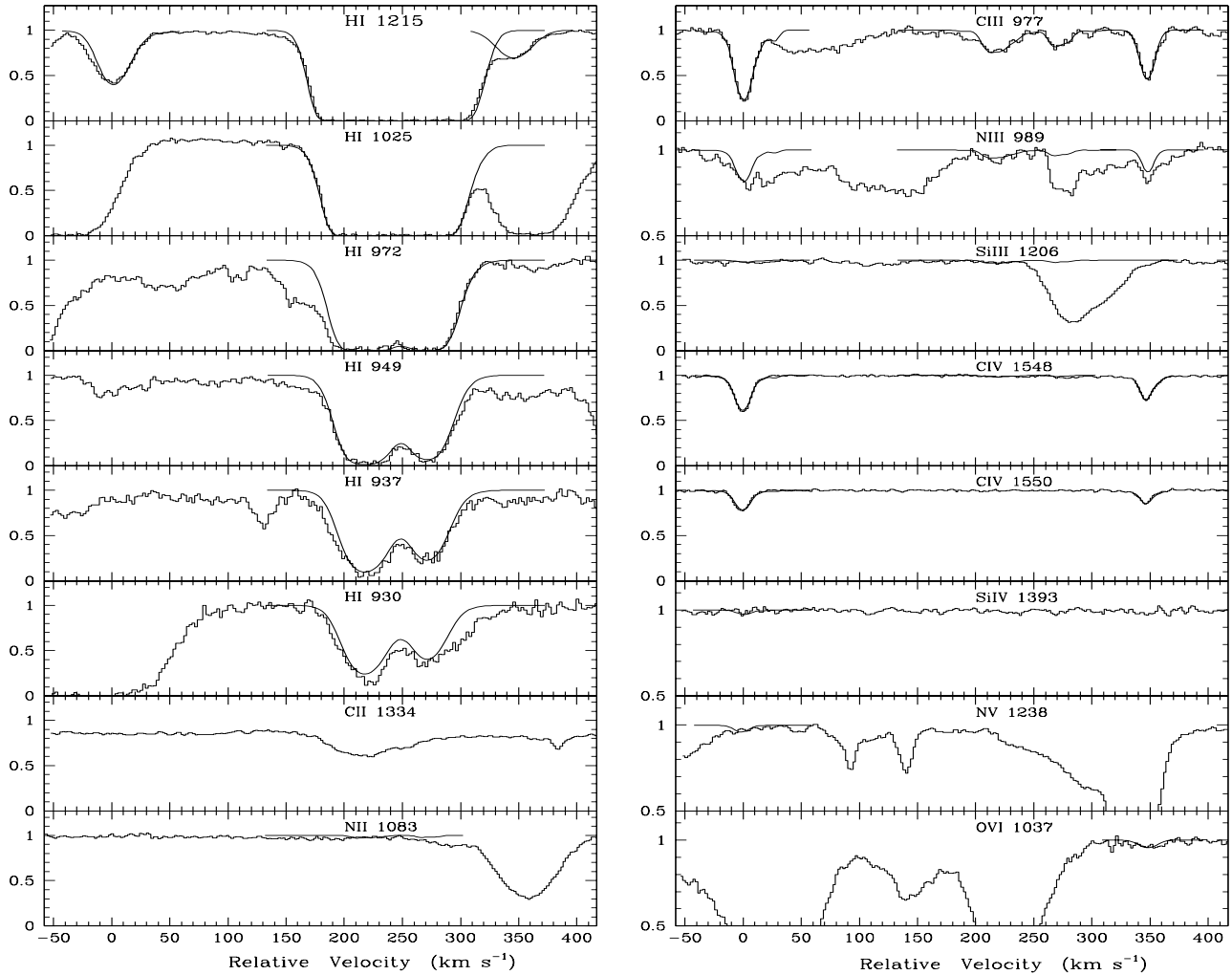,height=14cm,width=18cm}
\vspace{0.0cm}
\caption[]{Same as  Fig.~\ref{fg_1} but for
the complex absorption system at \zabs = 2.4158 towards \object{HE0151--4326}.
Three subcomponents cover the ranges:
$-30 < v < 40$ \kms\ ($A$),  $140 < v < 340$ \kms\ ($B$), and
$300 < v < 380$ \kms\ ($C$).
}
\label{fg_8}
\end{figure*}

\begin{figure*}[t]
\vspace{0.0cm}
\hspace{-0.2cm}\psfig{figure=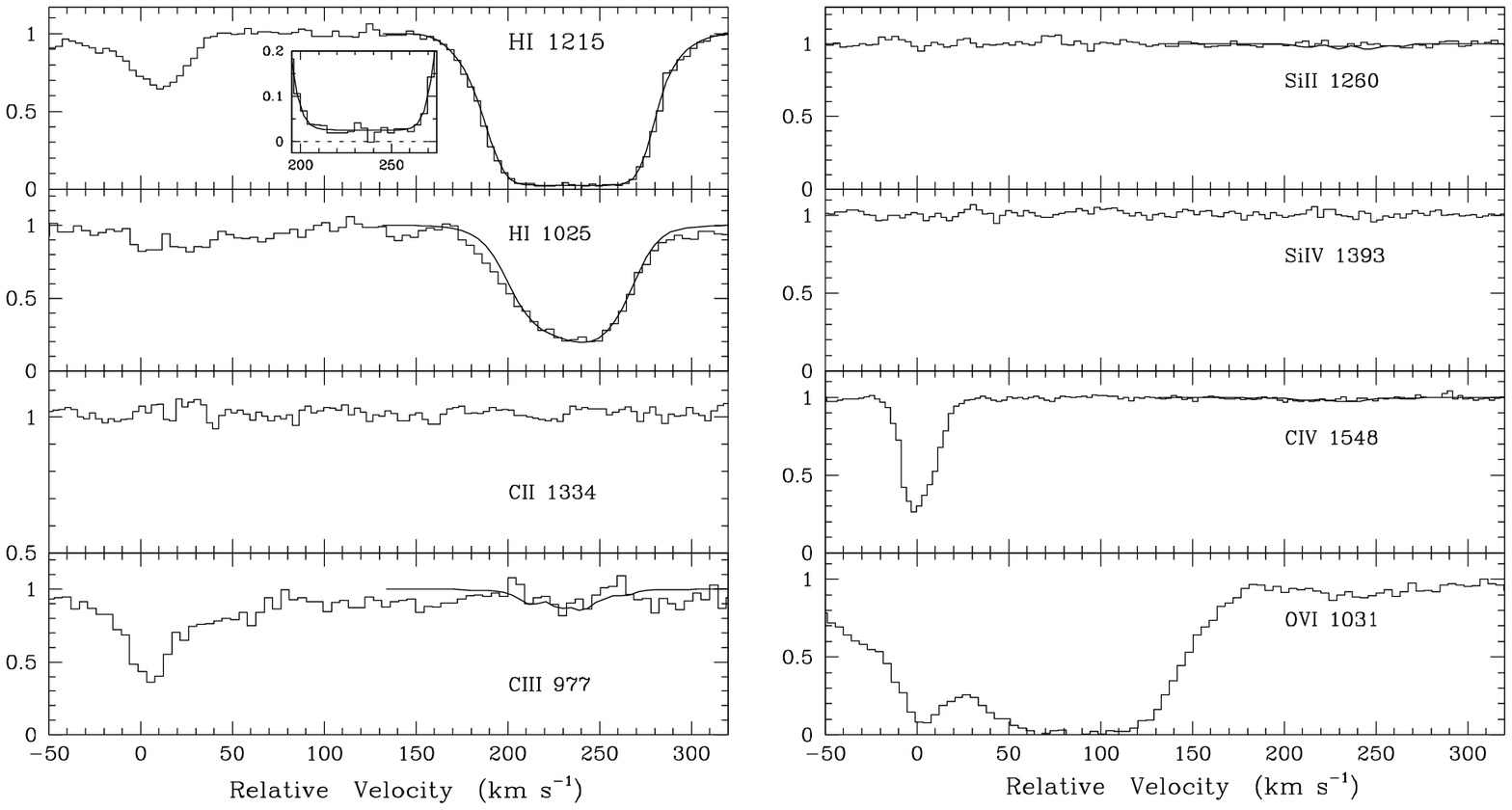,height=14cm,width=18cm}
\vspace{-6.5cm}
\caption[]{Same as  Fig.~\ref{fg_1} but for
the complex absorption system at \zabs = 2.3520 towards \object{Q0329--385}.
The subsystem at $v =0$ \kms\ is described in Paper~I, Sect.~2.4.
The subsystem at $v =230$ \kms\ has \zabs = 2.3545.
}
\label{fg_8a}
\end{figure*}

In the subsystem $B$, 
the available limits \ion{C}{iv}/\ion{C}{iii} $< 0.15$ 
and \ion{N}{ii}/\ion{N}{iii} $< 0.6$ provide  bounds  
to the acceptable $U$ range: -3.2 $< \log(U) <$ -2.25. 
In this $U$ range, 
$\log(\Upsilon_{\scriptstyle \rm H\,\scriptscriptstyle I}) \sim$ 
-2.2~-- -3.2,
the fractions of \ion{C}{iii} and \ion{N}{iii}
remain almost constant, 
$\log(\Upsilon_{\scriptstyle \rm C\,\scriptscriptstyle III}) \approx
\log(\Upsilon_{\scriptstyle \rm N\,\scriptscriptstyle III}) \sim$ -0.05, 
and the fraction of \ion{Si}{iii} changes 
from -0.1 to -0.2. 
All values depend only weakly on the spectral shape. 
Thus, the measured column
densities lead to the following abundances: 
[C/H] $\sim$ -2.5~-- -1.5,  [Si/C] $<$ -0.5, and 
[N/C] $\sim$ 0.5~-- 0.6
(providing that
\ion{N}{iii} $\lambda989$ absorption in subsystem $B$ is real).

This kind of the abundance pattern when a low 
carbon content (1/300 to 1/30 solar) is accompanied by 
a strong underabundance of silicon (and probably by 3-times overabundance
of nitrogen)  
is highly unusual for a stand-alone intergalactic
absorber with a comparable neutral hydrogen column density, 
$N$(\ion{H}{i}) $\sim 10^{16}$ \cm, 
but it coincides with [Si/C] (and [N/C]) measured in 
the subsystems $A$ and $C$.  
Such a similarity points  
to a physical connection between all three absorbers.  
Note that the ionization parameter $U_0$ in the
metal-poor subsystem $B$ is signifucantly 
lower than that in the metal-rich subsystems $A$ 
and $C$, i.e.  the gas density in the subsystem $B$ is higher 
(the kinetic temperature is higher as well because of low metallicity).
 
\begin{figure*}[t]
\vspace{0.0cm}
\hspace{-0.2cm}\psfig{figure=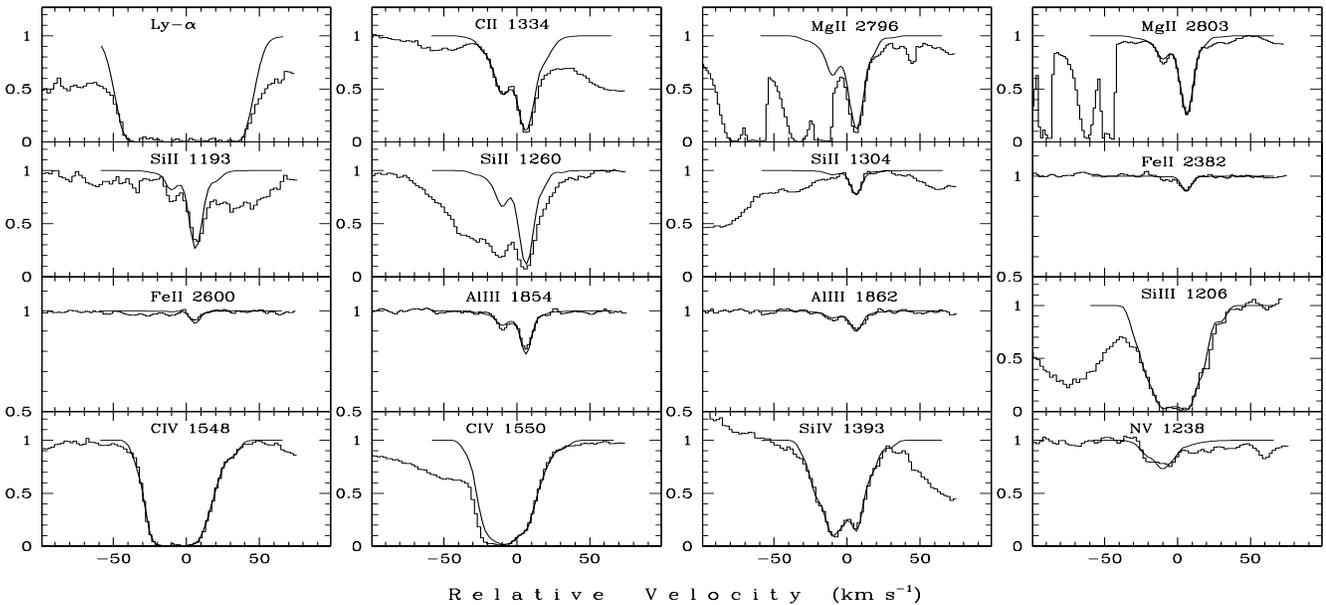,height=14cm,width=18cm}
\vspace{-5.7cm}
\caption[]{Same as Fig.~\ref{fg_1} but for
the \zabs = 1.7315 system (\object{HE0151--4326}).
}
\label{fg_9}
\end{figure*}

Absorption systems with strong metal lines and low
column densities of neutral hydrogen, 
$N$(\ion{H}{i}) $< 10^{14}$ \cm, such as the subsystems $A$
and $C$, are known from the literature (e.g., Schaye \etal\ 2007).
For these systems the derived  metal
abundances may be artificially boosted because of a possible overionization
of \ion{H}{i}  (see, e.g.,  Sect. 2.3.1 in Paper~I). 
However, the upper limit on $N$(\ion{O}{vi}) $< 4.5\times10^{12}$ \cm and
the low value of the ratio $N$(\ion{O}{vi})/$N$(\ion{C}{iv}) $< 0.65$ 
in the subsystem $C$ show that gas is close to the ionization equilibrium.
Thus, the measured high metal abundances are probably real. 

The obtained abundance pattern with silicon deficient 
(and nitrogen overabundant) 
to carbon, resembles patterns 
(taking silicon as a proxy to oxygen) observed
in planetary nebulae in the Milky Way as well as in the LMC and SMC 
(Stanghellini 2007).
Carbon (and nitrogen) are obviously enhanced due to dredge-up processes in the
progenitor star(s). Whether a low content of silicon 
is intrinsic or affected by some
depletion into dust is not clear since there are no other elements available.

It is known that many planetary nebulae/AGB-stars have cold outer envelopes
(see, e.g., Marengo 2009, and references therein).
In order to explain systematically higher abundances derived from weak 
optical recombination lines as compared to the
abundances resulting from collisionally excited lines, 
a model of planetary nebula (PN) 
assuming the presence of H-deficient (i.e. metal-rich)
inclusions embedded in the diffuse nebular gas is discussed
(e.g., Middlemass 1990; Liu \etal\ 2004; Wesson \& Liu 2004; Zhang \etal\ 2005).
Accounting for this model, the \zabs = 2.4158 absorption system
can be interpreted as PN fragment(s) transported into the
IGM by the AGN/galactic wind.

In Paper I, we described a system at \zabs = 2.3520 
towards \object{Q0329--385} with an oversolar
carbon content and overabundance of nitrogen to carbon 
(Table~\ref{tbl-3}). This system
exhibits a weak \ion{H}{i} Ly-$\alpha$ accompanied by strong metal lines 
(\ion{C}{iv}, \ion{N}{v}, \ion{O}{vi}), i.e.
resembles the subsystems $A$ and $C$, but is much higher ionized. 
The \zabs = 2.3520 system is shifted by $\sim$7400 \kms\ from
\object{Q0329--385} and was classified in Paper~I
as a probable eject from the quasar host galaxy. 
While analysing this system, we have not considered the
nearby absorption systems. However, detached 
by only 230 \kms\ (\zabs = 2.3545) 
there is an absorber with properties similar to
those of the subsystem $B$: strong \ion{H}{i} Ly-$\alpha$ 
along with clear continuum at the positions of the
\ion{C}{ii}, \ion{C}{iv}, \ion{Si}{ii}, 
\ion{Si}{iv}, and \ion{N}{v} lines (Fig.~\ref{fg_8a}). 
Weak absorptions are present at the
positions of \ion{C}{iii} $\lambda977$ and 
\ion{O}{vi} $\lambda1032$, but since both fall in the Ly-$\alpha$ forest, 
the observed intensities are only upper limits for these lines. 
The column density of \ion{H}{i} can be estimated quite
accurately (Ly-$\beta$ and Ly-$\gamma$ available): 
$N$(\ion{H}{i}) = $(4.5\pm0.3)\times10^{14}$ \cm. 
With such a high column density Ly-$\alpha$ should
be fully saturated~-- but it shows a flat bottom with the
residual intensity of 0.025
(see insert in Fig.~\ref{fg_8a}).
This means that the absorber is small and does not completely  
cover the light source. 
At the same time,  
the system at \zabs = 2.3520 with $N$(\ion{H}{i}) $\sim 10^{13}$ \cm, 
the total hydrogen column $N$(H) of a few
units of $10^{17}$ \cm\ and the gas density $n > 10^{-4}$ \cmm\ 
(Table~\ref{tbl-1})
does not show incomplete coverage, i.e. in any case
it is not smaller than the system at \zabs = 2.3545.
Therefore, the absorber with $N$(\ion{H}{i}) = $4.5\times10^{14}$ \cm\ 
should have
$N$(H) $< 10^{18}$ \cm\ and $n > 10^{-3}$ \cmm\ which is realized at 
$\log U < -2$, i.e. it 
cannot be a highly ionized system (and apparent absorption at the position of 
\ion{O}{vi} $\lambda1032$ is obviously due to some forest line).
With the upper limit on $N$(\ion{C}{iii}) $< 3.6\times10^{12}$ \cm\
we obtain for the carbon content the estimate [C/H] $< -1.0$.
Unfortunately, positions of \ion{Si}{iii} $\lambda1206$ and 
\ion{N}{iii} $\lambda$989 are blended and any conclusive
estimates on silicon and nitrogen contents are impossible. 
Thus, the system at \zabs = 2.3545 comprises 
low-ionization  and metal-poor gas~--- as was the case for the subsystem $B$,
whereas the \zabs = 2.3520 absorber~--- just as the subsystems $A$ and $C$~--- 
is formed by metal-rich and highly ionized gas. Surprisingly, even the
velocity offset between the \zabs = 2.3520 and \zabs = 2.3545 systems exactly
corresponds to that between the subsystems $A$ and $B$.

\subsubsection{System at \zabs = 1.7315}
\label{sect-2-2-2}

This system is very similar to that at \zabs = 1.7529 towards 
\object{HE1347--2457} (Sect.~\ref{sect-2-1-2}). 
A narrow saturated \ion{H}{i} $\lambda1215$ line is blended in 
the blue and red wings 
by hydrogen lines from the Ly-$\alpha$ forest (Fig.~\ref{fg_9}). 
Both the \ion{Si}{iv} lines 1393 \AA\ and 1402 \AA\
coincide with broad intrinsic \ion{O}{vi} absorption
(\object{HE0151--4326} is a mini-BAL quasar), 
but the profile of \ion{Si}{iv} $\lambda1393$ can be deconvolved. 
The line \ion{Mg}{ii} $\lambda2796$ is contaminated by telluric absorption. 
The lines of \ion{C}{iv} $\lambda\lambda1548, 1550$ fall in the 
Ly-$\alpha$ forest, \ion{C}{iv} $\lambda1550$ is blended in the blue
wing. 
Thus it is not clear whether the incomplete coverage 
is realized or not.
In the following all lines are considered as having covering factors of unity.
Acceptable shapes of the ionizing spectra are shown in Fig.~\ref{fg_10}, 
the derived physical parameters, column densities and the corresponding 
element abundances are presented in Tables~\ref{tbl-1}, \ref{tbl-2}, 
and \ref{tbl-3}.

Silicon is considerably depleted relative to carbon, whereas the magnesium 
to carbon ratio is almost two times higher than the solar value.
There is also  deficit of iron with respect to carbon. Similar
pattern characterized by the strong overabundance of magnesium and 
deficit of iron (deficit of silicon was not reliably detected), 
was found  in the absorption
system at \zabs = 1.7963 towards \object{HE2347--4342} (\#~10 in 
Table~\ref{tbl-1}, see also Sect. 3.1.2 in Agafonova \etal\ 2007). 
The overabundance of magnesium is observed
in AGB-stars and is related to the so-called `hot-bottom' 
burning (Herwig 2005).
  
Again, as in the case of the systems at \zabs = 1.7529 and \zabs = 1.5080
towards \object{HE1347--2457} (Sect.~\ref{sect-2-1-2}  and \ref{sect-2-1-3},
respectively), the shift of the magnesium lines to $\sim 0.6$ \kms\
leads to a noticeable improvement of the fitting. 
The present system shows a narrow but blended
line at the position of \ion{N}{v} $\lambda1238$.
Since the expected position of \ion{N}{v} $\lambda1242$
is blended with a strong forest absorption, the column density obtained from the
\ion{N}{v} $\lambda1238$ line is in fact an upper limit.
A conservative limit on the nitrogen
abundance shows that the overabundance of nitrogen
to carbon does not exceed 0.15 dex (1.4 times).

\begin{figure}[t]
\vspace{-1.0cm}
\hspace{-1.5cm}\psfig{figure=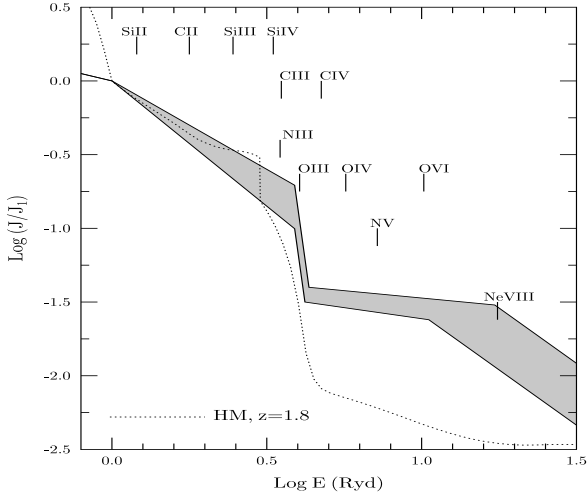,height=11cm,width=11cm}
\vspace{-3.5cm}
\caption[]{Same as Fig.~\ref{fg_2} but for
the \zabs = 1.7315 system (\object{HE0151--4326}). 
}
\label{fg_10}
\end{figure}

\subsection{Quasar \object{HE0001--2340} }
\label{sect-2-3}

\subsubsection{System at \zabs = 1.6514}
\label{sect-2-3-1}

This system consists of a narrow hydrogen \ion{H}{i} $\lambda1215$ line, 
saturated and blended in the blue and red
wings, and many strong lines of metal ions in different ionization 
stages (Fig.~\ref{fg_11}).
The simultaneous
fitting of all lines assuming constant metallicity throughout the absorber
makes it possible to restore the wings of the hydrogen Ly-$\alpha$ line. 
The line of \ion{C}{ii} $\lambda1334$ is
blended with a weak forest absorption and can be deconvolved. 
The expected positions of
the \ion{N}{v} $\lambda\lambda1238, 1242$ lines are contaminated with 
a strong intervening absorption
which prevents setting any reasonable limit on the nitrogen abundance. 
The shift of the
\ion{Mg}{ii} $\lambda2796, 2803$ lines by
$\sim 0.6$ \kms\ results in the improved fitting of
all low-ionization lines, \ion{C}{ii} $\lambda1334$,
\ion{Si}{ii} $\lambda1526$, and \ion{Mg}{ii} $\lambda\lambda2796, 2803$
(see Sect.~\ref{sect-2-1-2}).
It cannot be excluded that the magnesium
profiles are affected by an incomplete coverage, 
${\cal C}$(\ion{Mg}{ii}) $\sim 0.95-0.98$, but the current
quality of the spectral data does not allow us to make unambiguous conclusion.
The physical parameters are given in Table~\ref{tbl-1}. 
The column densities listed in Table~\ref{tbl-2} 
correspond to ${\cal C} = 1$ for all lines.  

Very similar ratios \ion{C}{iv}/\ion{C}{ii} = 
\ion{Si}{iv}/\ion{Si}{ii} = \ion{Si}{iii}/\ion{Si}{iv} $\sim 1.1$
require a hard ionizing spectrum at $1 < E < 4$ Ryd with 
$\alpha \sim 0.8-1$ 
and with the intensity break at
$E = 4$ Ryd of about 1 dex (Fig.~\ref{fg_12}). 
The resulting element abundances are given in Table~\ref{tbl-3}. 
Noticeable is an extremely strong depletion of silicon to carbon
and magnesium, [Si/C,Mg] $\sim -0.9$. 
The line \ion{Fe}{ii} $\lambda2382$ falls in the spectral region
with many weak telluric lines, so it is not clear whether the absorption at the
expected position of this lines is indeed due to \ion{Fe}{ii}
(no other iron lines are available). In any case, even an upper limit 
on the iron content shows that iron
is depleted as well, [Fe/C] $< -0.3$. 
In contrast to all systems described above the
present one has a deficit of aluminium: [Al/C] $< -0.6$. 
A continuum window at
the expected position of the \ion{O}{i} $\lambda1302$ line allows us
to set a very conservative limit on the oxygen
abundance, [O/H] $< 0.5$, which excludes any overabundance of oxygen to carbon.   
The linear size along the line-of-sight is small with an upper limit 
(the intergalactic $J_{912}$ intensity is assumed) of $L < 30$ pc.

\subsubsection{System at \zabs = 1.5770}
\label{sect-2-3-2}

There are three systems~--- at \zabs = 1.5770, 1.5810, and 
1.5855~---
located quite closely in the radial velocity range 
($\Delta v \sim 1000$ \kms)  and showing a similar
set of strong and narrow lines of different metal ions.
However, in the two latter systems most lines are blended, so the 
quantitative analysis is possible only for the former absorber
which is shown in Fig.~\ref{fg_13}.

\begin{figure*}[t]
\vspace{0.0cm}
\hspace{-0.2cm}\psfig{figure=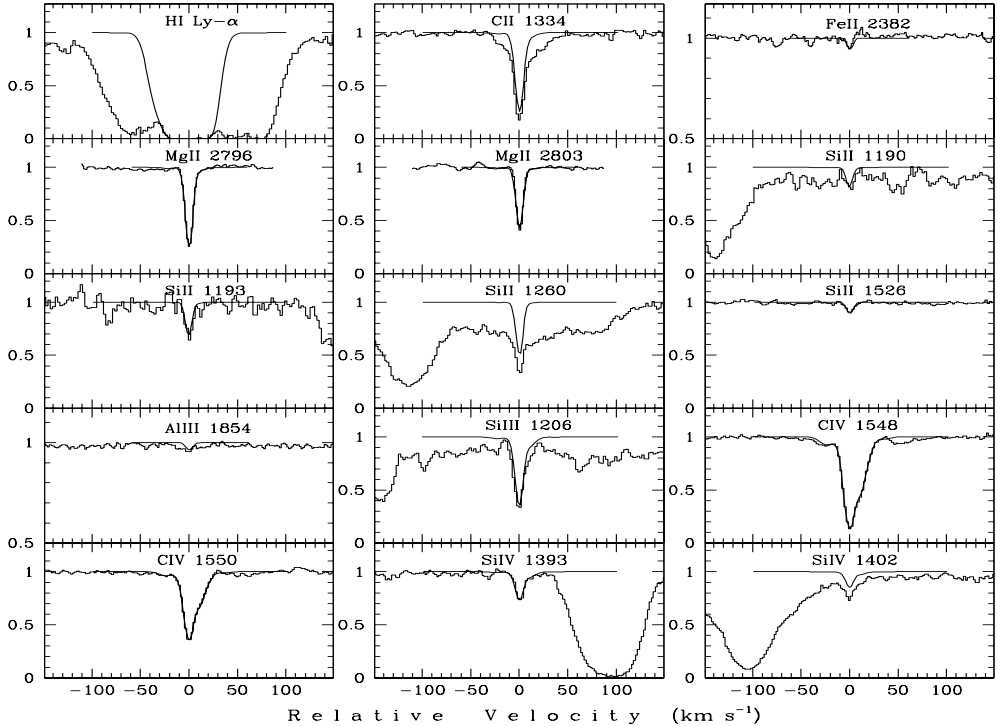,height=14cm,width=18cm}
\vspace{-4.0cm}
\caption[]{Same as Fig.~\ref{fg_1} but for
the \zabs = 1.65136 system (\object{HE0001--2340}). 
}
\label{fg_11}
\end{figure*}

From silicon lines, only \ion{Si}{iii} $\lambda$1206 and 
\ion{Si}{iv} $\lambda$1393 are present, \ion{Si}{iv} $\lambda$1402
is blended with deep forest absorption. Both lines fall in the Ly-$\alpha$ 
forest range and strictly speaking should be considered as upper limits (although
no metal contaminants were found). A narrow absorption at the position of
\ion{N}{v} $\lambda$1238 is very probably due to this ion 
(no metal candidates for blending),
its red wing is blended with shallow forest line, but can be deconvolved.
\ion{Mg}{ii} $\lambda$2796 coinsides with a telluric line, weak absorption 
at the position of \ion{Mg}{ii} $\lambda$2803 can also be telluric. 

For any type of the tried ionizing spectra the carbon content comes 
out extremely high~--- above 10 solar values. 
The apparent profile of Ly-$\alpha$ is inconsistent with the assumption of
constant metallicity throughout the absorber 
(see synthetic profiles in Fig.~\ref{fg_13}).
Note that similar inconsistency between
profiles of hydrogen lines and lines of metals was also detected
in Paper~I for two absorbers with very high metallicities:
the system at \zabs = 2.898 towards 
\object{HE2347--4243}, and the system at
\zabs = 2.352 towards \object{Q0329--385}. Both of them are associated.

The physical parameters and column densities are given in 
Tables~\ref{tbl-1}-\ref{tbl-3}.  
At such high metallicities as obtained for the \zabs = 1.5770 system
the dependence of ion fractions on the absolute metal abundances (which
are unknown) overrides the dependence on the SED, so it is not possible to 
bound the acceptable range of spectral shapes. 
However, UVB spectra which are softer at $E > 4$ Ryd than the model spectrum
for $z \sim 2$ of Haardt\& Madau (1996) should be probably excluded since at 
$U$ corresponding
to the observed ratio \ion{C}{ii}/\ion{C}{iv} they do not
reproduce the observed upper limits on \ion{Si}{iii} and 
\ion{Si}{iv} and give an extremely large
(more than one magnitude) overabundance of nitrogen to carbon.
The previously described system
at \zabs = 1.6514 for which the spectral shape can be restored is
detached in the velocity space by $\sim$8000 \kms. 
In principle, such velocity dispersions could be possible if 
absorbers were ejected by a quasar located transverse to the 
line-of-sight.  
In any case, for all types of trial spectra
silicon remains considerably underabundant to carbon, 
[Si/C] $< -0.7$. 
Also for all spectra, the deconvolved
\ion{N}{v} $\lambda1238$ line leads to overabundance of nitrogen to
carbon. The value of [N/C] depends on the softness at $E >$ 4 Ryd: the
softer the UVB spectrum the higher [N/C]. 

Extreme high abundances of carbon and nitrogen might be artificial because
of non-equilibrium ionization (\ion{H}{i} overionized) in the system considered.
However, more plausible explanation seems to be a
really H-deficient gas from the inner parts
of the envelopes of the AGB-stars. 
The low ratio of silicon to carbon and nitrogen, 
[Si/C,N] $< -1$, can be explained by
the enhancement of C and N due to the dredge-up processes 
and by the depletion of
silicon into dust.

\section{ Discussion }
\label{sect-3}

\subsection{Abundance patterns of high metallicity gas}
\label{sect-3-1}

In the present paper we considered 7 optically-thin systems 
with solar to oversolar metallicities. 
Thus, together with previously described absorbers, total
sample of metal-rich systems consists of 11 absorbers with
redshifts $1.5 < z < 2.9$ identified in spectra of 6 quasars. 
To simplify the reading, in the following the systems will be referred 
to by their numbers as given in Tables~\ref{tbl-1} and \ref{tbl-3}. 
Metallicity is indicated on base of the measured abundance of carbon.

The majority of the systems except \#~9 
(Tables~\ref{tbl-2}, \ref{tbl-3}) show 
abundance patterns which~--- in spite of being different~--- 
can nevertheless be related to outflows from
low and intermediate mass stars ($1 M_\odot < M < 10 M_\odot$)
in their post-main sequence evolution stages. 

The observed diversity of patterns supports the assumption that most metal-rich
absorbers have very small sizes ($L \sim 10-500$ pc) as compared 
with the intergalactic clouds,
i.e. absorbers are formed by outflows from only a few stars 
(or even from a single one).

The system \#~9  shows extreme high 
metal content exceeding 10 solar values.
It is not clear whether the apparent metallicity
is artificially boosted 
due to the overionized state of \ion{H}{i} or it is intrinsic and thus indicating
really hydrogen-deficient gas (see Sect. 2.3.1 in Paper~I). 
In any case, the overabundance of oxygen to carbon points
to SNe~II as the main source of gas enrichment. 
The presence of only one SNe~II-enriched system in our 
sample is probably a selection effect:
in quasar spectra there is a non-negligible population 
of absorbers revealing strong \ion{C}{iv} lines along with weak 
\ion{H}{i} absorption (see, e.g., Schaye \etal\ 2007). 
These systems may indeed contain high 
metallicity gas enriched prevalently by SNe~II,
but an accurate quantitative analysis is in most cases 
prevented by the absence of lines in subsequent ionization stages.

Fig.~\ref{fg_15} shows the ratio [N/C] as a function of the
carbon abundance [C/H] which in the present
case serves as an indicator of the overall metallicity.  
No positive correlation of [N/C] with
metallicity is seen: at any value of [C/H] nitrogen 
can be either over- or underabundant.
Similar findings were reported for planetary nebulae 
(Aller \& Czyzak 1983; Perinotto 1991; Richer \& McCall 2008) 
and for high-metallicity gas near AGN (Fields \etal\ 2005). 
This indicates that the nitrogen enrichment 
occurs quite irregular and is probably governed by many processes. 
On the other hand, the spread of the nitrogen contents
can be considered as an additional argument 
in favor of a very small linear size of the metal-rich
absorbers (formed by outflows from only a few stars).

\begin{figure}[t]
\vspace{0.0cm}
\hspace{-0.2cm}\psfig{figure=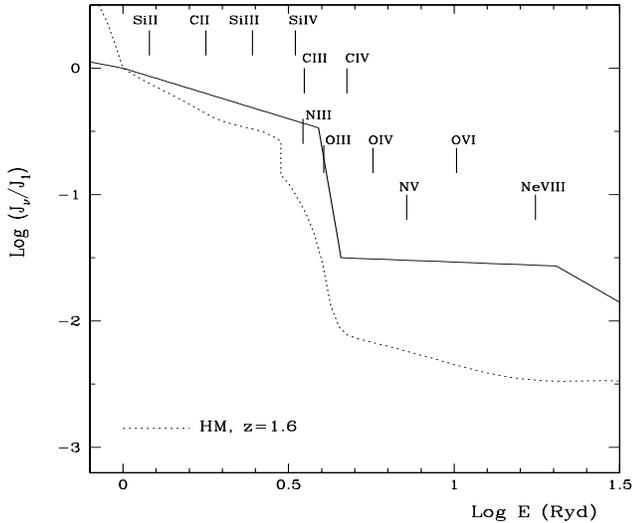,height=8.5cm,width=10.5cm}
\vspace{-1.5cm}
\caption[]{
The ionizing spectrum (solid line) corresponding to the ionization 
state observed in the \zabs = 1.65136 system (\object{HE0001--2340}).
The intergalactic ionizing spectrum at $z = 1.6$ modeled by   
Haardt \& Madau (1996) is shown by  the dotted line.
The spectra are normalized as in  Fig.~\ref{fg_2}.
}
\label{fg_12}
\end{figure}

Absorbers \#~1, 8, 9 and 11 belong to the so-called associated,
or proximate systems, 
i.e. they are located close to the background quasar and are formed
in gas ejected from the host galaxy. All other systems are
formally intervening. However, they demonstrate same physical properties
as the above associated~--- compare, for example,  \#~1 and 2,
\#~8 and 3, \#~11 and 4.
Two conclusions follow. Firstly, many metal-rich intervening systems
originate in gas expelled from the quasar host galaxies located
transverse to the line of sight. This is also supported by the direct
observations of \object{HE2347--4342} where transverse
quasars were detected just at the redshifts of such systems 
(Worseck \etal\ 2007).
Secondly, associated
systems not nessesary originate in gas ejected from the circumnuclear
regions~--- they can be interstellar objects (fragments of star/PN envelopes,
diffuse clouds etc.) entrained by the quasar wind and transported into
the intergalactic space. However, metal abundances obtained 
from the associated systems
are often used to verify models of quasar chemical evolution (e.g. 
D'Odorico \etal\ 2004; Fechner \& Richter 2009). 
Obviously, before making this,
the nature of a particular associated system should be clarified.

\begin{figure*}[t]
\vspace{0.0cm}
\hspace{-0.2cm}\psfig{figure=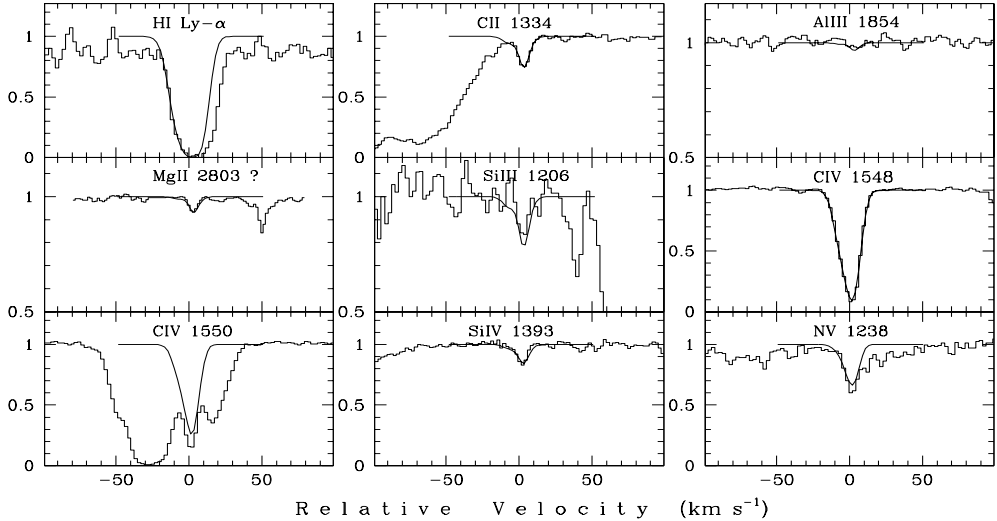,height=16cm,width=18cm}
\vspace{-8.5cm}
\caption[]{
Same as Fig.~\ref{fg_1} but for
the \zabs = 1.5770 system (\object{HE0001--2340}). 
Identification of the \ion{Mg}{ii} $\lambda2803$ line
is putative since the stronger component \ion{Mg}{ii} $\lambda2796$
is blended with telluric absorption. 
}
\label{fg_13}
\end{figure*}

\subsection{Depletion of silicon into dust}
\label{sect-3-2}

In Sect. \ref{sect-2-1-1} we argued that the deficit 
of silicon to carbon obtained for the metal-rich gas may be due
to condensation of silicon into dust in the envelopes of post-main
sequence stars and the subsequent gas-dust separation
in the stellar/galactic wind. Here we consider some aspects of
silicon depletion into dust in more detail.
 
In the oxygen-rich environment, C/O $< 1$,
the main dust species are iron and silicates consisting of olivine
Mg$_{2x}$Fe$_{2(1-x)}$SiO$_4$, pyroxen
Mg$_x$Fe$_{1-x}$SiO$_3$ and quartz SiO$_2$ 
(Gail \& Sedlmayr 1999; Jeong \etal\ 2003; Ferrarotti \& Gail 2006),
with olivine and pyroxen observed mostly in iron-free (i.e. $x = 1$) forms
(Jaeger \etal\ 1998; Waters \etal\ 1998; Draine 2003;
Gutenkunst \etal\ 2008; Perea-Calder\'on \etal\ 2009). 
Although processes of dust growth are far from being clear 
in full detail, all reactions related to the silicate dust formation 
require the presence of the molecule SiO.  This molecule is indeed 
observed in the envelopes of oxygen-rich AGB- and RSG-stars 
(van Loon \etal\ 2008; Sloan \etal\ 2008). 
Contrary to the previous standpoint
that SiO polimerizes at low temperature $\sim$600 K 
and therefore cannot form so-called seed nuclei (Jeong \etal\ 2003),
recent results have shown that SiO nucleates in fact 
at significantly higher temperature
$T \sim 1200$ K and forms small SiO grains 
over which the other dust components start to condense 
at continuously lower $T$ (Nuth \& Fergusson 2006).
Thus, at the initial stages of dust formation the most depleted
elements are expected to be oxygen and silicon, with  magnesium 
and iron following at the later stages. Observational evidences of 
strong magnesium gradients throughout some planetary nebulae
along with more or less uniform depletion of silicon can be considered as a
support for this picture
(P\'equignot \& Stasi\'nska 1980; Harrington \& Marionni 1981; Middlemass 1988).

We note that systems with [Si/C] $< 0$ 
(\#~2, 5, 6 in Table~\ref{tbl-3}) do not show deficit of silicon to iron 
which is  not in line with usual assumption that  
iron should be stronger depleted into dust than silicon. 
In principle, it is possible
that the depletion of iron is stronger in our systems as well: the intrinsic
iron content is unknown and it can be enhanced relative to other
elements as was obtained, e.g., for the absorbers \#~3 and 8 (and is observed in
giant stars in local dwarf galaxies~--- see Sect. \ref{sect-2-1-3}). 
However, the growth of dust grains
includes many stages and is strongly affected by the environmental conditions such
as gas metallicity, temperature, star pulsations etc.
Under some conditions iron becomes more depleted than other elements, 
under other~--- does not.  
Model calculations of the dust growth processes in AGB-stars by 
Ferrarotti \& Gail (2006) show that the relative depletion of Si and Fe 
depends on the metallicity and initial stellar mass, with
silicate dust prevalent over iron dust in 
high-metallicity (solar to oversolar) AGB-stars
with masses $M < 2 M_\odot$ or in massive AGB-stars $M \ga 4M_\odot$ 
experiencing hot-bottom burning.
The hot-bottom burning is considered also as the cause 
of the enhanced content of heavy
magnesium isotopes $^{25}$Mg and $^{26}$Mg (Herwig 2005) 
indications of which were detected
in the absorbers \#~2, 3, 5, and 6 
(Sect. \ref{sect-2-1-2}, \ref{sect-2-1-3}, 
\ref{sect-2-2-2}, \ref{sect-2-3-1}, and \ref{sect-2-3-2}, respectively).  
Thus, we may conclude that the absorbers in question are formed mainly from gas
expelled from massive AGB-stars. 

Note in passing that among metal-rich DLAs and LLSs, there are
objects with both low [Si/Fe]: [Zn] = 0.28, [Si/Fe] $< -0.5$ 
(P\'eroux \etal\ 2008),
[Zn] = 0.25, [Si/Fe] $< -0.4$ (Meiring \etal\ 2007), 
and high [Si/Fe]: [Zn] $> 0.86$, [Si/Fe] = 0.75
(Meiring \etal\ 2008), [Zn] = 0.12, [Si/Fe] = 0.7 (Prochaska \etal\ 2006).
Additionally, many DLAs shows differential depletion into dust with values
[Si/Fe] varying an order of magnitude from one component to another (Centuri\'on
\etal\ 2003; P\'eroux \etal\ 2006; Quast \etal\ 2008).

\subsection{Evolutional effects in distribution of metal-rich absorbers}
\label{sect-3-3}

On base of the observed lines, the metal-rich absorbers can be divided into
two distinct groups: systems revealing strong lines of both low 
(\ion{C}{ii}, \ion{Mg}{ii}, \ion{Si}{ii}) and
high (\ion{C}{iii}, \ion{N}{iii}, \ion{Al}{iii},
\ion{Si}{iii}, \ion{C}{iv}, \ion{Si}{iv}) ionization species 
(\# 1, 2, 3, 5, 6, 8, 10) 
and systems without or with weak low-ionization lines (\# 4, 7, 9, 11). 
The first group has the mean ionization parameter
$\log U < -2$, whereas the second one is characterized by 
$\log U > -1.5$.  
In Fig.~\ref{fg_16}, the $U$-values are plotted
against the redshifts of the systems for which they are obtained.  
The absence of systems from the second group at $z < 2.3$ in our sample 
is explained by pure selection effects:
the important for these systems lines of  \ion{C}{iii} 
$\lambda977$, \ion{N}{iii} $\lambda989$, and
\ion{O}{vi} $\lambda\lambda1032, 1037$  are shifted at $z < 2.3$
to the unobservable or very noisy wavelength range, 
and the remaining \ion{H}{i} $\lambda1215$, 
\ion{C}{iv} $\lambda\lambda1548, 1550$,
and eventually 
\ion{N}{v} $\lambda\lambda1238, 1242$, and 
\ion{Si}{iv} $\lambda\lambda1393, 1402$ 
do not allow to fix the ionization parameter and, hence, to derive 
accurate element contents. 

As for low-ionization systems, we find a single system
at $z > 2$ (\#~1 with \zabs = 2.5745)
and six systems at $z < 2$.   
In contrast to the high-ionization group, this result is not
affected either by  selection criteria for absorbers or by the
wavelength coverage of available quasar spectra 
(low redshift boundary \zabs = 1.5
is determined by the optical limit 
to observe  \ion{H}{i} $\lambda1215$) and, thus, reflects  a real
distribution of  low-$U$ absorbers over redshift. 

The epoch $z \sim 2$ is known to represent a peak activity in both
the cosmic star formation rate and the assembly of galaxies and it is also
characterized by the maximum space density of the luminous quasars
(Fan \etal\ 2001, Dickinson \etal\ 2003, Chapman \etal\ 2005,
Rudnick \etal\ 2006). 
Galaxies at $z \sim 2$ reveal star formation rates of 
$10-300 M_\odot$ yr$^{-1}$
and stellar masses $M_\ast \sim 10^9 - 10^{11.5} M_\odot$
(Genzel \etal\ 2008, and references cited therein).
The estimates of the comoving number density of the star-forming
galaxies from different surveys (Chen \etal\ 2003, Adelberger \etal\ 2005,
Sumiyoshi \etal\ 2009, Crist\'obal-Hornillos \etal\ 2009)
provide a density $n_g$ of a few $10^{-3} h^3$ Mpc$^{-3}$. 

Taking into account the detected jump in the number of low-ionization
absorbers at \zabs $< 2$, it is interesting  
to compare the space density of
these absorbers to the space density of galactic population at $z \sim 2$. 

At $z = 2$, the \ion{H}{i} Ly-$\alpha$ is shifted to 3647 \AA.\
Among the studied spectra of 20 QSOs, only half of them cover 
the wavelength range $\lambda < 3647$ \AA,
thus allowing us to identify low-ionization absorbers at $z < 2$.
In spectra with the coverage $\lambda < 3647$ \AA,
the probability to find at least one low-ionization metal-rich absorber
clearly exceeds 50\% . 
Assuming that the mean linear size of the absorber is $\sim 50$ pc,
the chance that a random sightline passes through a metal-rich cloud 
requires the comoving number density of the
absorbers $n_a \sim 1/L^2D$. With a typical distance lag $D \sim 10$ Gpc
($z \sim 1.8$),
the observed absorber frequency is realized if the comoving
density of these systems is $n_a \sim 4\times10^4$ Mpc$^{-3}$. 
Here
the comoving density of star-forming galaxies is taken as
$n_g \approx 2.4\times10^{-3}$ Mpc$^{-3}$ ($h = 0.7$)
which is in line with Chen \etal\ (2003).
Then there are 
$2\times10^7$ metal-rich absorbers in the vicinity of each star-forming galaxy
at $z \la 2$. 
This estimate is quite rough because of
the unknown geometric parameters of the clouds 
and the uncertainties in the galactic counting statistics. 
However,  it corresponds by an order of magnitude to
recent results on the Milky Way halo
where a population of more than $10^8$ weak metal-line absorbers
located in circumgalactic environment was detected (Richter \etal\ 2009). 
Richter \etal\ select their systems on base of the
presence of a weak \ion{O}{i} $\lambda$1302 line 
which supposes the ionization parameter $U$
of about $\log U \sim -3$ (cf, our system \#~10). 
The physical parameters of these Milky Way `gas wisps' are very much alike
to their high-redshift counterparts:
the hydrogen column density $N$(\ion{H}{i}) ranges from $2\times10^{16}$ \cm\
to $3\times10^{18}$ \cm, 
the hydrogen volume density
$n_{\scriptscriptstyle \rm H} \sim 0.02-0.3$ \cmm\ (if clouds are located at
the distance 50 kpc from the Galactic center), the absorber thicknesses are
$L \approx 0.2-130$ pc, 
and the metallicities are $\sim 0.1-1$ solar.

\begin{figure}[t]
\vspace{-0.5cm}
\hspace{0.0cm}\psfig{figure=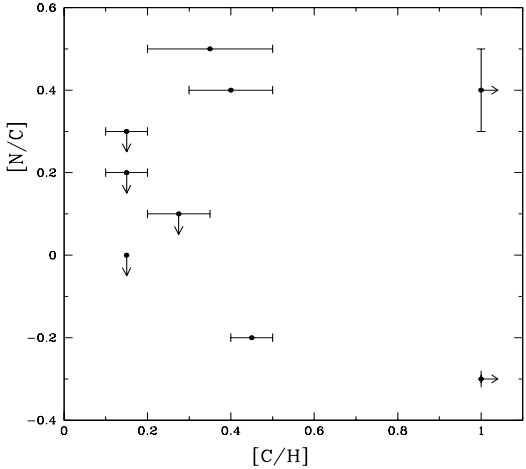,height=9cm,width=10cm}
\vspace{-2.5cm}
\caption[]{Relative ratio
[N/C] versus metallicity index [C/H] for metal-rich absorbers
(see Table~\ref{tbl-3}).
}
\label{fg_15}
\end{figure}

Finally, to estimate the mass of the clouds,
we assume that they are homogeneous spheres of radius $R$. Then,
$$
M_a = \frac{4}{3}\pi R^3 n_{\scriptscriptstyle \rm H} 
\mu m_{\scriptscriptstyle \rm H}
\approx 0.1\mu\left(\frac{n_{\scriptscriptstyle \rm H}}{{\rm cm}^{-3}}\right)
\left(\frac{R}{\rm pc}\right)^3 M_\odot\, ,
$$
where $\mu$ is the mean molecular weight of the gas, and 
$m_{\scriptscriptstyle \rm H}$ is the mass of the hydrogen atom.

For ionized gas of solar metallicity ($X:Y:Z = 0.710:0.265:0.025$ by mass), 
$\mu = 0.62$. With $n_{\scriptscriptstyle \rm H} \sim 10^{-2}$ \cmm\
and $R \sim 25$ pc, we obtain for a single cloud a typical mass of 
$M_a \sim 10 M_\odot$, i.e. low-ionization metal-rich absorbers can indeed
be produced by only a few intermediate-mass stars.
For a star-forming galaxy with
$M_\ast \sim 2\times10^{10} M_\odot$, the total mass $M_{\rm tot}$ of all
metal-rich absorbers reaches
$M_{\rm tot} \sim 0.01 M_\ast$. 

The peak star forming activity at $z \sim 2$ is 
usually explained by the presence
of large amounts of gas supplied either by mergers or by smooth gas accretion
onto galaxies along the large-scale filaments, with probable prevalence of
the second mechanism 
(Genel \etal\ 2008; Shapiro \etal\ 2008; Genzel \etal\ 2008). 
Our data show
that galaxies at $z \la 1.8$ have already undergone stellar evolution sufficient
to provide high (solar to oversolar) level of metallicity for the
large population of stars. The build-up of metals requires many
generations of stars. Estimations from SINFONI
survey of galaxies at $z \sim 2$ show that star-forming
discs experienced constant star formation rates over at least 0.5 Gyr
(Daddi \etal\ 2007). This  means more than ten
generations for intermediate mass stars~--- probably enough time to acquire
high metallicity.
  
\begin{figure}[t]
\vspace{-0.5cm}
\hspace{0.0cm}\psfig{figure=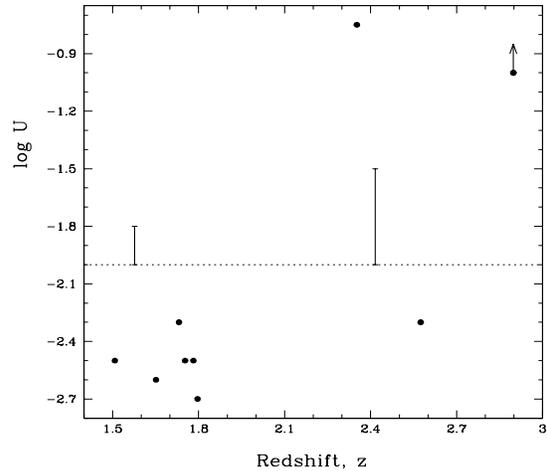,height=9cm,width=10cm}
\vspace{-2.5cm}
\caption[]{Ionization parameter
$\log U$ estimated from metal-rich absorbers with redshifts 
$1.5 < z < 2.9$ (see Table~\ref{tbl-2}). The dashed line
separates absorbers with low and high $U$-values.
}
\label{fg_16}
\end{figure}

The high metal content has positive
effect on the star mass-loss rates  through driving the outflow both by 
radiative acceleration
in resonance lines of metal ions and by radiation pressure on the dust, 
the formation of which
is in turn boosted in the metal-rich environment. Thus, at high metallicity the
recycling of matter occurs more rapid leading to constantly large amount of
gas present in the interstellar space. On the other hand, starbursts
and possible AGNs in the central regions favor the
formation of the large-scale winds which expel gas from galaxies. 
At some redshift,
more gas is expelled as returned resulting in decline of both star forming
and AGN activities. This point occurs somewhere below redshift 1.5. 
Unfortunately,
wavelength coverage of optical telescopes limits redshifts
of absorption
systems suitable for quantitative analysis~--- below $z = 1.5$ 
the hydrogen \ion{H}{i} $\lambda1215$ line 
is not seen and, hence, accurate metallicity of the absorbing gas 
cannot be determined.
Commissioning of the Cosmic Origins Spectrograph~---
the high-resolution UV spectrograph with wavelength coverage from 1100 to 3200 
\AA~---
and the World Space Observatory (WSO)-UV will change the situation allowing to
probe high-metallicity systems at redshifts below 1.5.

\section{ Summary and conclusions }
\label{sect-4}

We report on the analysis of 11 metal-rich absorption systems with redshifts
$1.5 < z < 2.9$ identified in spectra of 6 distant quasars. 
Due to a special selection of
the absorption systems and the advanced methods to solve
the inverse spectroscopy problem the obtained abundance patterns can be 
considered as reliable. On base of these patterns, we identify types of stars
responsible for the enrichment of the absorbing gas. The main results are
as follows.
\begin{enumerate} 
\item[1.]
The majority of the described metal-rich absorbers (10 from 11) 
reveal abundance patterns 
which~--- in spite of being different~--- can be attributed 
to the outflows from low and intermediate
mass stars (LIMS) in their post-main
sequence evolutional stages (red giant branch and asymptotic giant branch). 
One absorber is enriched prevalently by SNe~II. 
A small number of SNe~II-enriched 
objects does not represent real statistics,
but is a consequence of selection criteria. 
Diversity of observed patterns
points to small linear sizes of the metal-rich absorbers, i.e. every cloud 
can be formed by outflows from only a few (or even from a single) star(s). 
\item[2.] 
Several absorbers show a significant deficit of 
silicon with respect to other elements. 
This deficit can be explained by the depletion of Si into 
dust in the oxygen-rich envelopes of the post-main sequence stars 
and the subsequent dust-gas separation in the galactic/quasar wind.
\item[3.]
If lines of \ion{Mg}{ii} $\lambda\lambda2796, 2803$ 
are centered using the laboratory wavelengths which correspond 
to the solar isotopic ratio
$^{24}$Mg:$^{25}$Mg:$^{26}$Mg = 79:10:11, 
then profiles of all low-ionization lines 
(\ion{C}{ii}, \ion{Mg}{ii}, \ion{Si}{ii}, \ion{Fe}{ii})
cannot be fitted self-consistently with a required 
$\chi^2_{\rm min} \sim 1$.  
The quality of fitting improves significantly if centering of the
\ion{Mg}{ii} lines occurs using slightly shorter wavelengths which 
correspond to an enhanced content of heavy
isotopes $^{25}$Mg and $^{26}$Mg in the absorbing gas. 
Although the quality of the available quasar spectra does
not allow to estimate accurate isotopic composition of magnesium, 
the enhancement of heavy Mg isotopes in the systems 
at \zabs = 1.7529, 1.7315, 1.6514,  and  1.5080 
seems to be very probable. Such an enhancement is predicted
for AGB-stars with masses $M \ga 4 M_\odot$ 
which experience a so-called hot-bottom burning.
\item[4.]
The abundance of nitrogen does not show correlation with metallicity: 
at any value of [C/H] nitrogen can be
both over- and underabundant to carbon. 
This means that the nitrogen enrichment occurs quite irregularly and
is governed by several processes.
\item[5.]
Among metal-rich systems, a group characterized by
low ionization parameter, $\log U < -2.3$, 
can be singled out. At $z > 2$, we find only one low-$U$ system
(\zabs = 2.5745), whereas at $z < 2$ six such systems are detected.
This result is not affected by any selection criteria and
reflects the real redhsift distribution of the low-ionization
metal rich systems. 
The comparison of the number densities 
of metal-rich absorbers and 
star-forming galaxies at $z \sim 2$ shows that there should 
be $\sim 10^7-10^8$ such absorbers around each
galaxy. This coincides with the number of small absorbing clouds 
detected recently around the Milky Way.
The redshift $z \sim 2$ is characterized by peak values of 
the star-formation rates and the quasar luminosity
function which are currently explained by the presence of large 
amounts of gas in the interstellar space
of high-redshift galaxies due to mergers and rapid gas accretion along the
large-scale filaments, with the second option more preferable by 
recent observations of the star-forming galaxies at $z \sim 2$. 
A high metallicity can be another cause for the peak activities 
at this redshift: large
amount of metals in the envelopes of LIMS boosts the mass-loss 
from stars both due to radiative acceleration in resonance lines of metal 
ions and by radiation pressure on the dust grains. This leads
to a rapid recycling of matter and, hence, to a constantly 
high amount of gas present in the interstellar space. 
When the amount of gas expelled from a galaxy due to the large-scale wind 
exceeds the amount of gas supplied by accretion and stellar winds, the
star-forming and AGN activities decline. 
In order to study these processes through the analysis of
metal-rich absorbers~---  
which turned out to be quite reliable indicators of cosmic 
evolution~--- joint optical/UV observations of quasar spectra are needed.
This is expected to occur with future Cosmic Origine Spectrograph and WSO-UV.
\end{enumerate}

 \begin{acknowledgements}
S.A.L. and I.I.A. gratefully acknowledge the hospitality 
of Osservatorio Astronomico di Trieste, 
Hamburger Sternwarte, and 
the Shanghai Astronomical Observatory
while visiting there. 
This research has been supported by 
the RFBR grant No. 09-02-00352, 
and by 
the Federal Agency for Science and Innovations grant
NSh 2600.2008.2.
J.L.H. is supported by the National Science Foundation of China
No. 10573028, the Key Project No. 10833005, the Group Innovation Project
No. 10821302, and by the 973 program with No. 2007CB815402.
\end{acknowledgements}


\begin{thebibliography}{}
\bibitem{}Adelberger, K. L., Steidel, C. C., Pettini, M., \etal\ 2005,
ApJ, 619, 697

\bibitem{}Agafonova, I. I., Levshakov, S. A., Reimers, D., \etal\ 
2007, A\&A, 461, 893

\bibitem{}Agafonova, I. I., Centuri\'on, M., Levshakov, S. A., \& Molaro, P. 
2005, A\&A, 441, 9

\bibitem{}Aller, L. H., \& Czyzak, S. J. 1983, ApJS, 51, 211

\bibitem{}Aracil, B., Petitjean, P., Pichon, C., \& Bergeron, J. 2004, A\&A, 419, 811
 
\bibitem{}Asplund, M., Grevesse, N., \& Sauval, A. J. 2004, Nucl. Phys. A, 777, 1

\bibitem{}Bashkin, S., \& Stoner, J. O., Jr. 1975, {\it Atomic Energy
Levels and Grotrian Diagrams} (North-Holland Pub. Company: Amsterdam)

\bibitem{}Bonifacio, P., Sbordone, L., Marconi, \etal\ 2004, A\&A, 414, 503

\bibitem{} Calura, F., \& Matteucci, F. 2004, MNRAS, 350, 351

\bibitem{}Centuri\'on, M., Molaro, P., Vladilo, G., \etal\ 2003, A\&A, 403, 55
 
\bibitem{}Chapman, S. C., Blain, A. W., Smail, I., \& Ivison, R. J. 
2005, ApJ, 622, 722

\bibitem{}Chen, H.-W., Marzke, R. O., McCarthy, P. J., \etal\ 2003, ApJ, 586, 745

\bibitem{}Cohen, J. G., Huang, W., Udalski, A., Gould, A., \& Johnson, J. A.
2008, ApJ, 682, 1029

\bibitem{}Crist\'obal-Hornillos, D., Aguerri, J. A., L., Moles, M., 
\etal\ 2009, ApJ, 696, 1554

\bibitem{}Curdt, W., Brekke, P., Feldman, U., \etal\ 2001, A\&A, 375, 591

\bibitem{}Daddi, E., Dickinson, M., Morrison, G., \etal\ 2007, ApJ, 670, 156

\bibitem{}Davies, B., Origlia, L., Kudritzki, R.-P., \etal\ 2009, ApJ,  
696, 2014

\bibitem{}Dickinson, M., Papovich, C., Ferguson, H. C. \& 
Budav\'ari, T. 2003, ApJ, 587, 25

\bibitem{}D'Odorico, V., Cristiani, S., Donatella, R., Granato, G.L., \& 
Danese, L. 2004, MNRAS, 351, 976

\bibitem{}Draine, B. T. 2003, ARA\&A, 41, 241

\bibitem{}Fabbian, D., Recio-Blanco, A., Gratton, R. G., \& Piotto, G. 2005,
A\&A, 434, 235 

\bibitem{}Fan, X., Strauss, M. A., Schneider, D. P. \etal\ 2001, AJ, 121, 54

\bibitem{}Fardal, M. A., Giroux, M. L., \& Shull, M. 1998, AJ, 115, 2206

\bibitem{}Fechner, C., \& Richter, P. 2009, A\&A, 496, 31

\bibitem{}Ferland, G. J., Korista, K. T., Verner, D. A., et al. 1998, PASP,
110, 761 

\bibitem{}Ferrarotti, A. S., \& Gail, H.-P. 2006, A\&A, 447, 553

\bibitem{}Fields, D., Mathur, S., Pogge, R.W., et al. 2005, ApJ, 634, 928

\bibitem{}Gail, H.-P., \& Sedlmayr, E. 1999, A\&A, 347, 594

\bibitem{}Genel, S., Genzel, R., Bouch\'e, N. \etal\ 2008, ApJ, 688, 789

\bibitem{}Genzel, R., Burkert, A., Bouch\'e, N., \etal\ 2008, ApJ, 687, 59

\bibitem{}Gutenkunst, S., Bernard-Salas, J., Pottasch, S. R., Sloan, G. C., 
\& Houck, J. R. 2008, ApJ, 680, 1206

\bibitem{}Haardt, F., \& Madau, P. 1996, ApJ, 461, 20 

\bibitem{}Hamann, F., \& Ferland, G. 1999, ARA\&A, 37, 487

\bibitem{}Harrington, J. P., \& Marionni, P. A. 1981, 
in {\it The Universe at Ultraviolet Wavelengths: The First Two Yrs. of Intern. 
Ultraviolet Explorer}, (Maryland Univ., College Park), pp. 623--631 

\bibitem{}Herwig, F. 2005, ARA\&A, 43, 435

\bibitem{}Holweger, H. 2001, in {\it Solar and Galactic Composition}, 
ed. R. F. Wimmer-Schweingruber, AIP Conf.Proc., 598, 23

\bibitem{}Hou, J. L., Boissier, S., \& Prantzos, N. 2001, A\&A, 370, 23 

\bibitem{}Jaeger, C., Molster, F. J., Dorschner, J., \etal\ 1998, A\&A, 339, 904

\bibitem{}Jenkins, E. B., Bowen, D. V., Tripp, T. M., \& Sembach, K. R.
2005, ApJ, 623, 767

\bibitem{}Jeong, K. S., Winters, J. M., Le Bertre, T., \& Sedlmayr, E. 2003, A\&A, 407, 191

\bibitem{}Karakas, A. I., \& Lattanzio, J. C. 2003, PASA, 20, 279

\bibitem{}Kelly, R. L. 1987, J. Phys. Chem. Ref. Data, vol. 16, No.1

\bibitem{}Levshakov, S. A., Agafonova, I. I., Reimers, D., Hou, J. L., \&
Molaro, P. 2008, A\&A, 483, 19 [Paper I]

\bibitem{}Levshakov, S. A., Molaro, P., D'Odorico, S., \etal\
2007, A\&A, 446, 1077 

\bibitem{}Levshakov, S. A., Agafonova, I. I., Centuri\'on, M., \& Molaro P. 
2003, A\&A, 397, 851

\bibitem{}Levshakov, S. A., Agafonova, I. I., \& Kegel, W. H. 
2000, A\&A, 360, 833 [LAK]

\bibitem{}Liu, Y., Liu, X.-W., Barlow, M. J., \& Luo, S. G.
2004, MNRAS, 353, 1251

\bibitem{}Marengo, M. 2009, arXiv: astro-ph/0902.1536

\bibitem{}Meiring, J. D., Kulkarni, V. P., Lauroesch, J. T., \etal\ 2008, 
MNRAS, 384, 1015

\bibitem{}Meiring, J. D., Lauroesch, J. T., Kulkarni, V. P., \etal\ 2007, 
MNRAS, 376, 557

\bibitem{}Middlemass, D. 1990, MNRAS, 244, 294

\bibitem{}Middlemass, D. 1988, MNRAS, 231, 1025

\bibitem{}Molaro, P., Levshakov, S. A., Monai, S., \etal\ 2008, A\&A, 481, 559

\bibitem{}Morton, D. C. 2003, ApJS, 149, 205

\bibitem{}Morton, D. C. 1991, ApJS, 77, 119

\bibitem{}Nuth, J. A., III, \& Ferguson, F. T. 2006, ApJ, 649, 1178

\bibitem{}Perea-Calder\'on, J. V., Garc\'ia-Hern\'andez, D. A., 
Garc\'ia-Lario, P., Szczerba, R., \& Bobrowsky, M. 2009, A\&A, 495, L5

\bibitem{}Perinotto, M. 1991, ApJS, 76, 687

\bibitem{}P\'eroux, C., Meiring, J. D., Kulkarni, V. P. \etal\ 2008, 
MNRAS, 386, 2209

\bibitem{}P\'eroux, C., Meiring, J. D., Kulkarni, V. P. \etal\ 2006, 
MNRAS, 372, 369

\bibitem{}P\'equignot, D., \& Stasi\'nska, G. 1980, A\&A, 81, 121

\bibitem{}Polletta, M., Weedman, D., H\"onig, S. \etal\ 2008, ApJ, 675, 960

\bibitem{}Prochaska, J. X., O'Meara, J. M., Herbert-Fort, S., \etal\ 
2006, ApL, 648, L97

\bibitem{}Quast, R., Reimers, D., \& Baade, R. 2008, A\&A, 477, 443

\bibitem{}Reimers, D., Agafonova, I.I., Levshakov, S.A., \etal\, 
2006, A\&A, 449, 9

\bibitem{}Reimers, D., Janknecht, E., Fechner, C., \etal\ 2005, A\&A, 435, 17

\bibitem{}Reimers, D., K\"ohler, T., \& Wisotzki, L. 1996, A\&AS, 115, 235

\bibitem{}Rich, R. M., Origlia, L., \& Valenti, E. 2007, ApJ, 665, L119

\bibitem{}Richer, M. G., \& McCall, M. L. 2008, ApJ, 684, 1190

\bibitem{}Richter, P., Charlton, J. C., Fangano, A. P. M., 
Ben Bekhti, N., \& Masiero, J. R. 2009, ApJ, 695, 1631

\bibitem{}Rudnick, G., Labb\'e, I., F\"orster Schreiber, N. M., \etal\ 
2006, ApJ, 650, 624

\bibitem{}Sandlin, G. D., Bartoe, J.-D. F., Brueckner, G. E., Tousey, R., \&
Van Hoosier, M. E. 1986, ApJS, 61, 801

\bibitem{}Schaye, J., Carswell, R. F., \& Kim, T.-S. 2007, MNRAS, 379, 1169

\bibitem{}Shapiro, K. L., Genzel, R., F\"orster Schreiber, N. M.,
\etal\ 2008, ApJ, 682, 231 

\bibitem{}Silverman, J. D., Lamareille, F., Maier, C., \etal\ 2009, ApJ, 
696, 396

\bibitem{}Simcoe, R. A., Sargent, W. L., Rauch, M., \& Becker, G. 
2006, ApJ, 637, 648

\bibitem{}Simcoe, R. A., Sargent, W. L., \& Rauch, M. 2004, ApJ, 606, 92

\bibitem{}Sloan, G. C., Kraemer, K. E., Wood, P. R., \etal\ 2008, ApJ, 686, 1056

\bibitem{}Songalila, A., \& Cowie, L. L. 1996, AJ, 112, 335

\bibitem{}Stanghellini, L. 2007, in {\it Why Galaxies Care About AGB Stars: 
Their Importance as Actors and Probes}, eds. F. Kerschbaum, C. Charbonnel, 
and R. F. Wing (San Francisco: Astronomical Society of the Pacific), p.456

\bibitem{}Sumiyoshi, M., Totani, T., Oshige, S., \etal\ 2009, arXiv: 0902.2064

\bibitem{}Tautvai\v{s}ien\.e, G., Geisler, D., Wallerstein, G., \etal\
2007, AJ, 134, 2318

\bibitem{}van Loon, J. Th., Cohen, M., Oliveira, J. M., \etal\ 2008, 
A\&A, 487, 1055

\bibitem{}Venn, K. A., Irwin, M., Shetrone, M. D., \etal\ 2004, AJ, 128, 1177

\bibitem{}Vladilo, G. 2002, A\&A, 391, 407

\bibitem{}Vladilo, G., \& P\'eroux, C. 2005, A\&A, 444, 461

\bibitem{}Waters, L. B. F. M., Beintema, D. A., Zijlstra, A. A., \etal\ 
1998, A\&A, 331, L61

\bibitem{}Wessson, R., \& Liu, X.-W. 2004, MNRAS, 351, 1026

\bibitem{}Wisotzki, L., Christlieb, N., Bade, N., \etal\ 2000, A\&A, 358, 77

\bibitem{}Wisotzki, L., K\"ohler, T., Groote, D., \& Reimers, D. 1996, 
A\&AS, 115, 227

\bibitem{}Worseck, G., Fechner, C., Wisotzki, L., Dall'Aglio, A. 2007,
A\&A, 473, 805

\bibitem{}Yong, D., Aoki, W., \& Lambert, D. L. 2006, ApJ, 638, 1018

\bibitem{}Yong, D., Lambert, D. L., \& Ivans, I. I. 2003, ApJ, 599, 1357

\bibitem{}Zhang, Y., Liu, X.-W., Luo, S.-G., P\'equignot, D., \&
Barlow, M. J. 2005, A\&A, 442, 249

\bibitem{}Zoccali, M., Hill, V., Lecureur, A., \etal\ 2008, A\&A, 486, 177

\end{thebibliography}
\end{document}